\PassOptionsToPackage{dvipsnames}{xcolor}
\documentclass[sigconf]{acmart}

\usepackage{balance}
\usepackage{textcomp}
\usepackage{mathtools}   
\usepackage{epigraph}
\usepackage{subfig}
\setcopyright{rightsretained}
\def\Section {\S}

\newcommand{\squishlist}{
 \begin{list}{$\bullet$}
  { \setlength{\itemsep}{0pt}
     \setlength{\parsep}{3pt}
     \setlength{\topsep}{3pt}
     \setlength{\partopsep}{0pt}
     \setlength{\leftmargin}{1.5em}
     \setlength{\labelwidth}{1em}
     \setlength{\labelsep}{0.5em} } }

\newcommand{\squishlisttwo}{
 \begin{list}{$\bullet$}
  { \setlength{\itemsep}{0pt}
     \setlength{\parsep}{0pt}
    \setlength{\topsep}{0pt}
    \setlength{\partopsep}{0pt}
    \setlength{\leftmargin}{2em}
    \setlength{\labelwidth}{1.5em}
    \setlength{\labelsep}{0.5em} } }

\newcommand{\squishend}{
  \end{list}  }

\acmDOI{10.1145/3267809.3267821}
\acmISBN{978-1-4503-6011-1/18/10}
\acmConference[SoCC '18]{SoCC '18: ACM Symposium on Cloud Computing}{October 11--13, 2018}{Carlsbad, CA, USA}
\acmYear{2018}
\copyrightyear{2018}
\setcopyright{acmcopyright}
\acmPrice{15.00}

\begin{document}

\title[Cloud Index Tracking]{Cloud Index Tracking: Enabling Predictable Costs in Cloud Spot Markets}

\author{Supreeth Shastri}
\affiliation{
  \institution{UMass Amherst}
}
\email{shastri@umass.edu}

\author{David Irwin}
\affiliation{
  \institution{UMass Amherst}
}
\email{deirwin@umass.edu}

\begin{abstract}
Cloud spot markets rent VMs for a variable price that is typically much lower than the price of on-demand VMs, which makes them attractive for a wide range of large-scale applications.  However, applications that run on spot VMs suffer from cost uncertainty, since spot prices fluctuate, in part, based on supply, demand, or both. The difficulty in predicting spot prices affects users and applications: the former cannot effectively plan their IT expenditures, while the latter cannot infer the availability and performance of spot VMs, which are a function of their variable price.  Prior work attempts to address this uncertainty by modeling and predicting individual spot prices based on historical data.  However, a single model likely does not apply to different spot VMs, since they may have different levels of supply and demand. In addition, cloud providers may unilaterally change spot pricing algorithms, as EC2 has done multiple times, which can invalidate existing price models and prediction methods.  

To address the problem, we use properties of cloud infrastructure and workloads to show that prices become more stable and predictable as they are aggregated together. We leverage this observation to define an aggregate \emph{index price} for spot VMs that serves as a reference for what users should expect to pay.  We show that, even when the spot prices for individual VMs are volatile, the index price remains stable and predictable.  We then introduce \emph{cloud index tracking}: a migration policy that tracks the index price to ensure applications running on spot VMs incur a predictable cost by migrating to a new spot VM if the current VM's price significantly deviates from the index price.  We implement cloud index tracking on EC2, and show that it yields a predictable cost near that of the index price, but with much higher availability compared to prior work, which aggressively migrates to the lowest cost VM. 
\end{abstract}



\maketitle

\section{Introduction}
\label{sec:introduction}
To maintain the illusion that infinite resources are always available on-demand, public cloud providers provision their infrastructure for their expected peak demands.  As a result, a significant fraction (up to $40\%$ by a recent estimate~\cite{dc-utilization}) of cloud  resources are idle.  To recoup some of the capital and operational expenses of maintaining this surplus computing capacity, cloud providers now offer it in the form of transient servers~\cite{yank-ieee}, which they can revoke at any time.  A prominent example of transient servers are the Spot Instances offered by Amazon's Elastic Compute Cloud (EC2).  EC2 charges a variable spot price for spot VMs, which is determined in part based on variations in their long-term supply, and reserves the right to revoke them at any time to satisfy higher-priority requests, e.g., for on-demand or reserved VMs.

Since spot VMs are unreliable and may experience revocations, spot prices tend to be 50-90\% less than the price of on-demand VMs, which EC2 tries not to revoke. Due to their low price, spot VMs are highly attractive to large-scale applications. For example, Fermilab's Scientific Computing Division used spot VMs to dynamically scale up their computing capacity by 4$\times$ to accelerate the discovery of the Higgs-Boson~\cite{ec2-higgs-boson}. Similarly, machine learning and natural language processing researchers recently set the record for the largest high-performance cloud cluster by using 1.1 million vCPUs on spot VMs~\cite{ec2-clemson}.  However, while spot VMs offer significant potential for cost savings, the magnitude of these savings is not guaranteed, is based on future prices, which are uncertain, and could be negative if prices or revocation characteristics unexpectedly change. This lack of predictability presents both technical and policy challenges.

Prior work has addressed many of the technical challenges associated with gracefully handling spot revocations by tuning fault-tolerance mechanisms to mitigate their performance impact~\cite{spoton,flint,tr-spark,pado,exosphere,spotcheck}. However, applications that execute on spot VMs must also address \emph{cost uncertainty}, since there is no guarantee on how high or low the spot price will go.  While variable spot pricing has many advantages, including improving the supply/demand balance and ensuring resources are always accessible to users (if they are willing to pay for them), the resulting cost uncertainty presents new policy challenges.  For example, spot VMs make it difficult for enterprises to follow standard practices of allocating fixed budgets for acquiring a fixed amount of cloud resources.  Such administrative complexities may be one reason competing providers, including Google Compute Engine and Microsoft Azure, introduced a simpler fixed-price model for transient servers despite the advantages of variable pricing.   Reducing cost uncertainty requires applications to both accurately predict spot prices, and then leverage those predictions to adapt their execution to maintain a fixed cost, e.g., by delaying execution or migrating to other resources if prices change. 

Predicting spot prices is challenging for several reasons. Most importantly, users generally must predict many different prices. For example, in EC2, each type of VM in each availability zone (AZ)\footnote{An AZ is akin to a separate data center} of each region has its own dynamic spot price, resulting in roughly 7500 different dynamic spot prices across 44 AZs in 16 regions. By comparison, there are only roughly 6000 stocks listed across the New York Stock Exchange and the NASDAQ.  In addition, while a number of researchers~\cite{drafts,deconstruct,sigcomm-bid,infocom-bid,no-bid,predict3,predict2,predict1} and startups~\cite{batchly,spotinst,clusterk} have proposed techniques for modeling and predicting spot prices, there is no guarantee a one-size-fits-all model exists as prices are based on local supply/demand conditions that need not correlate across VM types, AZs, or regions. Further, spot prices in EC2 are not necessarily market-based, but instead determined by Amazon based on an internal pricing algorithm, which they often change, making prior price models and prediction methods obsolete.  For example, Amazon modified its pricing algorithm in late 2017 to decrease price volatility by altering prices based on longer-term changes in supply, rather than based on short-term supply and demand. While we discuss the broader implications of this change in \Section\ref{sec:discussion}, it invalidated prior price models and prediction methods.  

Rather than predict prices, recent work proposes a reactive approach that instead monitors spot prices and continuously migrates applications to the most cost-efficient VM, i.e., the VM offering the lowest cost per unit of resource utilized~\cite{hotspot}.  This approach encapsulates applications in containers to enable such stop-and-copy migrations.  EC2's recent adoption of fine-grained per-second billing allows this approach to exploit price inversions and arbitrage opportunities that are both small and brief.   However, the approach does not improve cost certainty or performance predictability, as the most cost-efficient spot VM may change frequently, e.g., every few minutes, resulting in frequent periods of unavailability while migrating to chase the lowest prices.  In addition, migrating to chase low prices incurs an upfront cost overhead due to the migration, which may not pay for itself over time if prices change (and could even possibly increase cost).  Second order effects may also increase price volatility if users adopt such an approach \emph{en masse}, and everyone starts aggressively chasing the lowest price.  This may be one reason behind the recent change in EC2's pricing algorithm. 

To address the problem, we observe that investors face a similar issue in financial markets when making investment decisions: since predicting individual stock prices is challenging, investors base their decisions, in part, on the characteristics of broader market indices, such as the Dow Jones Industrial Average, S\&P 500, and NASDAQ, which aggregate the price of many stocks. These indices typically serve as a reference point, or baseline, for investors to evaluate their portfolio's performance and make decisions. In addition, by aggregating the price of many stocks, these broader indices tend to be less volatile than any individual stock's price. As we show, spot prices exhibit similar high-level characteristics:  an aggregate spot price, or spot index, across many VMs is less volatile and more predictable than any single VM's spot price, enabling such a spot index to also serve as a useful benchmark for making decisions.  Importantly, as we discuss, these characteristics are general and independent of any particular pricing algorithm.  We leverage the insights above to define an aggregate index price for spot VMs and use it to develop \emph{cloud index tracking}: a container migration policy that tracks the cloud index price to ensure applications running on spot VMs incur a stable and predictable cost by migrating to a new spot VM if the current VM's price significantly deviates from the index price. In doing so, we make the following contributions. 

\noindent {\bf Cloud Index Definition}.  We analyze historical spot prices to show that as we aggregate prices they become more stable and predictable, and also discuss the underlying reasons for such a stable index price based on current cloud infrastructure and workload characteristics.  We then define a cloud index using this approach to serve as a reference point for what users should expect to pay.

\noindent {\bf Cloud Index Tracking Migration Policy}. We use our index price to implement cloud index tracking: a migration policy that, rather than chase low prices, only migrates applications to a new spot VM if its prices deviates significantly from the index.   We show that this migration results in a predictable price near that of the index price and a high availability.  We also compare with other migration policies to expose a tradeoff between cost and availability: chasing low prices reduces cost (and predictability) but results in more migrations, which increase overhead and reduces availability. 

\noindent {\bf Implementation and Evaluation}. We implement cloud index tracking  in Python on EC2 and release it as open source.\footnote{Available at \url{https://umass-sustainablecomputinglab.github.io/cloudIndex}}  We evaluate cloud index tracking relative to policies from prior work, and show that it enables more predictable costs and higher availability. 

\section{Background and Motivation}
\label{sec:background}

Renting VMs for a variable price that is a function, in part, of supply or demand is advantageous, since it can attract additional demand (and revenue) for unused resources.  Since cloud platforms incur large capital expenses to provision their infrastructure, they have a strong incentive to maximize its utilization and revenue. In addition, variable pricing also increases resource obtainability~\cite{economy-spot}, as users can always acquire resources if they are willing to pay a high enough price for them.  In contrast, under a fixed price model, a cloud platform may run out of resources under periods of high demand, requiring it to reject any additional resource requests, as shown in prior work~\cite{spotlight}.  Companies are often wary of relying entirely on cloud platforms for their computing infrastructure due to the risk that they will not be able to acquire resources when necessary, which can result in large monetary losses.  Thus, variable-priced spot VMs are a useful alternative for acquiring resources when on-demand VMs are unavailable, as users can always acquire spot VMs if they are willing to pay a high enough price. 

\begin{figure}[t]
\centering
\includegraphics[width=0.4\textwidth]{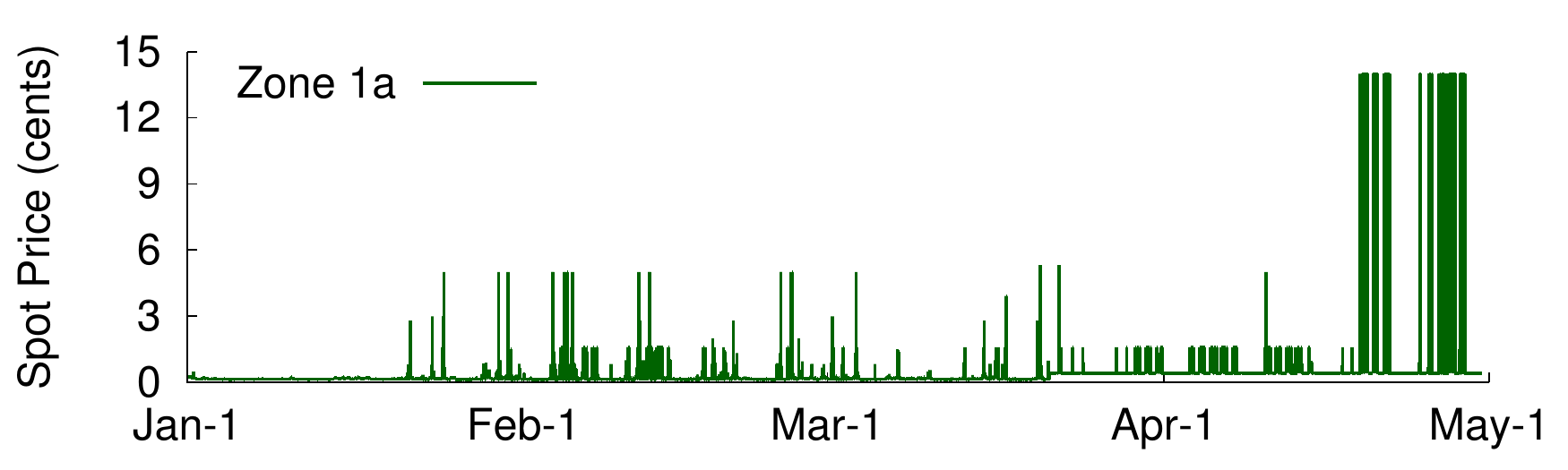}\\
\vspace{0.05cm}
\includegraphics[width=0.4\textwidth]{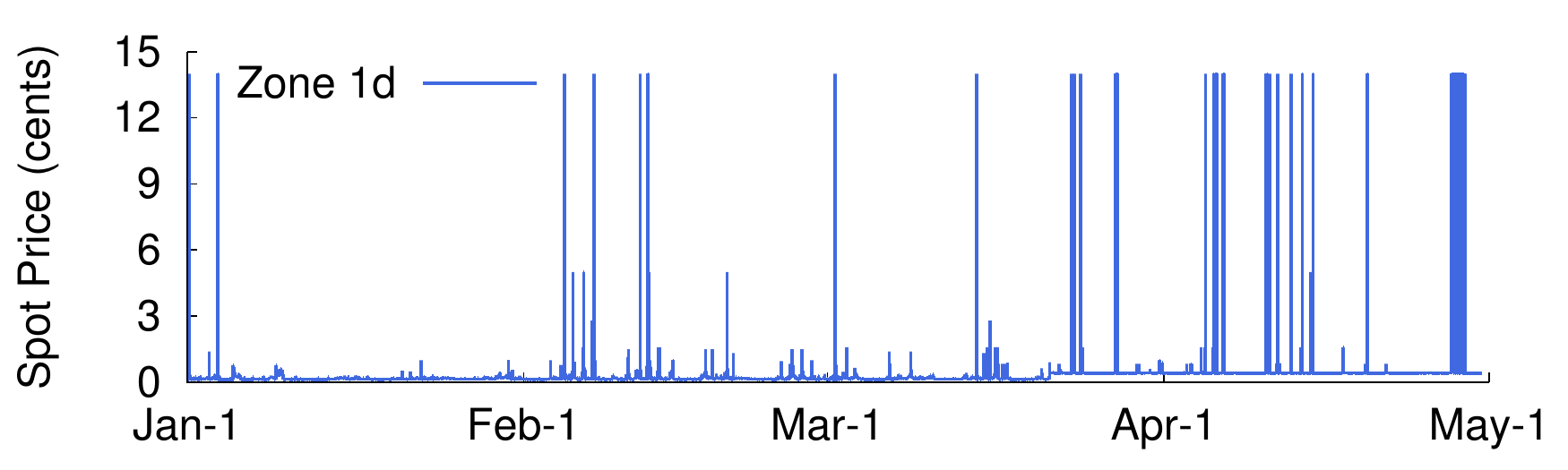}
\caption{Price of a representative Linux server ({\tt r3.4xlarge}) across two availability zones in us-east-1 region in 2017.}
\label{fig:individual}
\vspace{-0.3cm}
\end{figure}

However, variable pricing also has a number of drawbacks, largely due to its increased complexity.  For example, to enable accessibility for a high enough price, EC2 must be willing to revoke spot VMs from their current users and provision them to users willing to pay more.  Such revocations, which are akin to failures, incur an overhead that reduces the performance capacity (and thus value) of spot VMs~\cite{transient-guarantees,workshop2}. Many applications also either fail or perform poorly under unexpected and frequent revocations, and require significant modifications to gracefully handle them.  In addition, variable pricing also introduces \emph{cost uncertainty}, as there is no guarantee on the price of resources and thus an application's ultimate execution cost.  This is not amenable to many enterprises, which allocate fixed budgets for acquiring a fixed amount of resources.  

Unfortunately, current methods for predicting spot prices to estimate application cost are ineffective for a number of reasons. In particular, the sheer number of spot VMs with their own dynamic spot price makes prediction challenging. In EC2, there are currently 16 regions worldwide, each with 2-6 AZs, and some with over 50 types of VMs, resulting in more than 7500 spot prices to predict.  This scale will continue to increase as Amazon has plans to add 6 new regions and 17 more AZs in the near future~\cite{ec2-future}.  Each of these spot VMs has its own spot price dynamics, which reflect, in part, its own unique local supply and demand conditions.  To illustrate the challenge, Figure~\ref{fig:individual} shows that even the same spot VMs in different AZs of the same region may exhibit widely different spot price characteristics (in this case, a {\tt r3.4xlarge} Linux instance in four AZs of the {\tt us-east-1} region). Accurately predicting these different price dynamics for effectively the same resources, as well as the other 7500 spot VMs, without additional information is not feasible.   

In addition, even if accurate price predictions were possible, they would likely become obsolete once EC2 altered its pricing algorithm.  As prior research has noted~\cite{deconstruct}, EC2 is not required to set its spot prices based purely on supply and demand, since it both operates the market and owns all the resources.  In fact, EC2 has modified its spot pricing algorithm multiple times, most recently in late 2017, when it altered the pricing algorithm to reduce the magnitude and volatility of spot prices by tying them to longer-term changes in supply, rather than both supply and demand~\cite{ec2-pricing-algorithm}.  While we discuss the broader implications of this change in \Section\ref{sec:discussion}, it served to make prior spot price prediction methods invalid.  Importantly, the complexity and cost uncertainty of using spot VMs under variable pricing likely discourages users from adopting variable priced spot VMs, and may be one reason competitors have adopted simpler fixed-price models for similar types of revocable transient VMs.  

Recent work seeks to reduce this complexity by encapsulating cloud applications in resource containers and transparently migrating them to the most cost-efficient spot VM as prices change~\cite{hotspot}.  However, this approach of chasing the lowest price has some drawbacks.  In particular, chasing low prices does not actually reduce cost uncertainty:  since the lowest price may vary considerably, enterprises still cannot accurately predict the amount of spot resources they can purchase for a fixed budget.  In addition, repeated migrations to chase low prices can incur an unacceptable overhead for some applications.  For example, containers use stop-and-copy migrations, which result in downtime while their memory image is copied from one VM to another.  Prior work attempts to ensure migrations ``pay for themselves'' by only triggering them if a new spot VM is sufficiently cheaper than the current spot VM~\cite{hotspot}.  However, spot price changes can still cause a net loss for any migration. 

More importantly, many applications may perform poorly if VMs are frequently unavailable.  Examples of such applications include long-running applications that are occasionally interactive, such as data sinks for Internet-of-Things (IoT) devices and BitTorrent file trackers, as well as distributed applications that follow a bulk synchronous parallel (BSP) programming model implemented by many popular big data frameworks, such as Hadoop, Spark, and many others.  The former may perform poorly, since VMs may not be available when necessary to provide access to storage or data, while the latter may perform poorly, since all VMs running workers must synchronize at pre-defined barriers, causing frequent periods of unavailability to result in stragglers that delay other workers.   While these applications offer some flexibility to migrate to lower prices and restart after revocations, as they can tolerate occasional unavailable periods, the level of availability does affect their performance.  Prior work does not consider the application-level cost of such unavailability when making migration decisions. As we discuss, our approach to cloud index tracking considers both cost predictability and availability in determining when to migrate.

\subsection{Aggregate Spot Price Characteristics}
\label{sec:first-principles}

Prior work often analyzes historical spot VM prices, and models them by selecting a distribution that best characterizes the data.  As discussed earlier, such models generally do not apply across all spot VMs, and often become obsolete after changes to the pricing algorithm.  Thus, to improve cost predictability, we instead examine two underlying general properties of the spot prices based on common cloud infrastructure and workload characteristics.

\noindent {\bf Dependent Spot Prices.} Cloud platforms offer a wide range of VMs with different capacities to allow users to select the VM that best fits their application's workload characteristics.  While these different VMs have their own spot price, they are generally provisioned out of a fixed number of physical machines.  For example, EC2 offers 23 general-purpose VM types, e.g., the {\tt t2}, {\tt m3}, {\tt m4}, {\tt m5}, etc., that are likely hosted on only 4 different types of physical machines based on EC2's offering of only 4 dedicated non-virtualized hosts. Since the supply of physical machines does not significantly change over short timescales, e.g., hours or days, the spot prices of VMs within the same class are likely not completely independent.  That is, despite, say, the {\tt m4.large} and {\tt m4.16xlarge} having different spot prices, their price and revocation characteristics are likely partially correlated. Of course, many factors may affect the spot price, including administrative policies that determine how to allocate physical machine resources between VM types and demand for each type.  Prior work has largely ignored this underlying physical reality, and has often assumed that price and revocation characteristics are independent and identically distributed.   This potential dependent relationship between spot prices of different VMs further complicates modeling individual spot prices that share the same types of physical machines.   Aggregating the price of dependent VMs should mitigate these dependent effects.

\noindent {\bf Stable Idle Capacity}. In addition, while any VM's spot price may vary significantly, recent work and publicly-released traces of cloud data center resource utilization suggest that idle capacity is not only large, but also relatively stable~\cite{dc-utilization}.  For example, Microsoft recently detailed workload and utilization characteristics for Azure data centers~\cite{azure-economics} (along with publicly-released traces~\cite{resource-central}) that showed that, even though CPU utilization varies on the order of half the data center's capacity, most users do not dynamically scale their allocated capacity to match their utilization.  As a result, Azure does not experience large changes in resource allocations at the customer or data center level. Specifically, the reported median volatility for VM allocations was 6.3\% hourly, 2.6\% daily, and 3.2\% weekly at the data center level.  These results align with observations of Google data centers~\cite{economy-spot}, where researchers found large fractions of idle capacity are highly available (>98.9\%) over multi-month periods, which also indicates a relatively stable level of idle capacity. 

If compute capacity were i) an entirely fungible resource, such as other commodities, e.g., oil, electricity, corn, etc., ii) all of a data center's idle capacity were offered in a single market (for one per-unit price), and iii) application's were fully flexible, then a stable idle capacity implies that its clearing price would be largely stable and predictable (as both the supply and demand are stable).  While the assumptions above are not practical, a cloud index price, discussed below, offers a first order approximation that enables us to benefit from these observations and characteristics.

\subsection{Market Index}

A stable idle capacity implies that modeling prices in aggregate across many VMs, rather than for each individual VM, should yield more stable and predictable prices.  Two granularities are natural choices for aggregation: i) all spot prices of VMs within a family, which we assume are hosted on the same hardware resources, e.g., general-purpose, compute-optimized, and storage-optimized servers, and ii) all VMs within a data center (i.e., AZ in EC2) or region (composed of multiple AZs).  The former should eliminate dependencies related to sharing a common hardware base, while the latter should reflect the idle stable capacity we infer from prior work and publicly-available traces.  

Aggregating spot prices across many VMs is equivalent to computing a market index, which is a statistical measure of the value of a collection of items. Such market indices are commonly used in economics and finance to make decisions. For example, the Consumer Price Index (CPI) measures the changes in the price level of a pre-determined market basket of consumer goods purchased by typical households. Economists use the annual percentage change in CPI as a measure of inflation, which in turn guides a variety of economic policies. Similarly, stock market indices, such as the Dow Jones Industrial Average, the S\&P 500 and the NASDAQ, report the statistical measure of a prominent set of publicly traded stocks, and are considered to be broad indicators of the economy. 

As computation evolves into more of a commodity and an increasingly important investment, technology companies may also leverage such an index to succinctly describe the behavior of variable-priced spot VMs.  As in economics, an index provides a useful benchmark for quantifying resource cost, and, as we discuss next, is more stable and predictable than individual spot VM prices. 

\section{Cloud Index Definition}
\label{sec:cloud-index}

A cloud index succinctly quantifies the aggregate spot price characteristics of a large group of spot VMs. Below, we describe our approach to computing a cloud index, and then apply it to spot VMs in EC2 to characterize its salient features. 

\subsection{Cloud Index Definition}

Cloud platforms separate VMs into different types and set a different price based on both their CPU and memory capacity.  Network bandwidth and storage, including I/O bandwidth and space, are typically decoupled and separately billed.  In EC2, the compute capacity of VMs (and dedicated hosts) ranges from 1 to 349 EC2 Compute Units, or ECUs, which represent a relative measure of a CPU's integer processing power, while the memory capacity varies from 0.5GBs to nearly 2TB.   EC2 includes both compute-optimized and memory-optimized VMs which have higher CPU and memory capacity, respectively, relative to the other. Thus, one issue in defining a cloud index is computing a normalized per-unit price for CPU and memory resources to fairly compare the per-unit price of VMs with different resource allocations. Since the geometric mean is a common method for averaging different items, such as memory and CPU, with different numerical ranges, our cloud index uses it to compute a normalized price per unit of resource for each spot VM. Specifically, we define the normalized price $\hat{P}_i(t)$ of a spot VM $i$ at time $t$ per unit of CPU and memory resource as below, where $P_i(t)$ represents VM $i$'s spot price (in \$/hour) at time $t$, while $C_i$ and $M_i$ represent its CPU and memory capacity. 

\vspace{-0.2cm}
\begin{equation}
\label{eq:cost-efficiency}
\hat{P}_i(t) = \frac{P_i(t)}{\sqrt{C_i \cdot M_i}}
\end{equation}

\begin{figure}[t]
\centering
\includegraphics[width=0.45\textwidth]{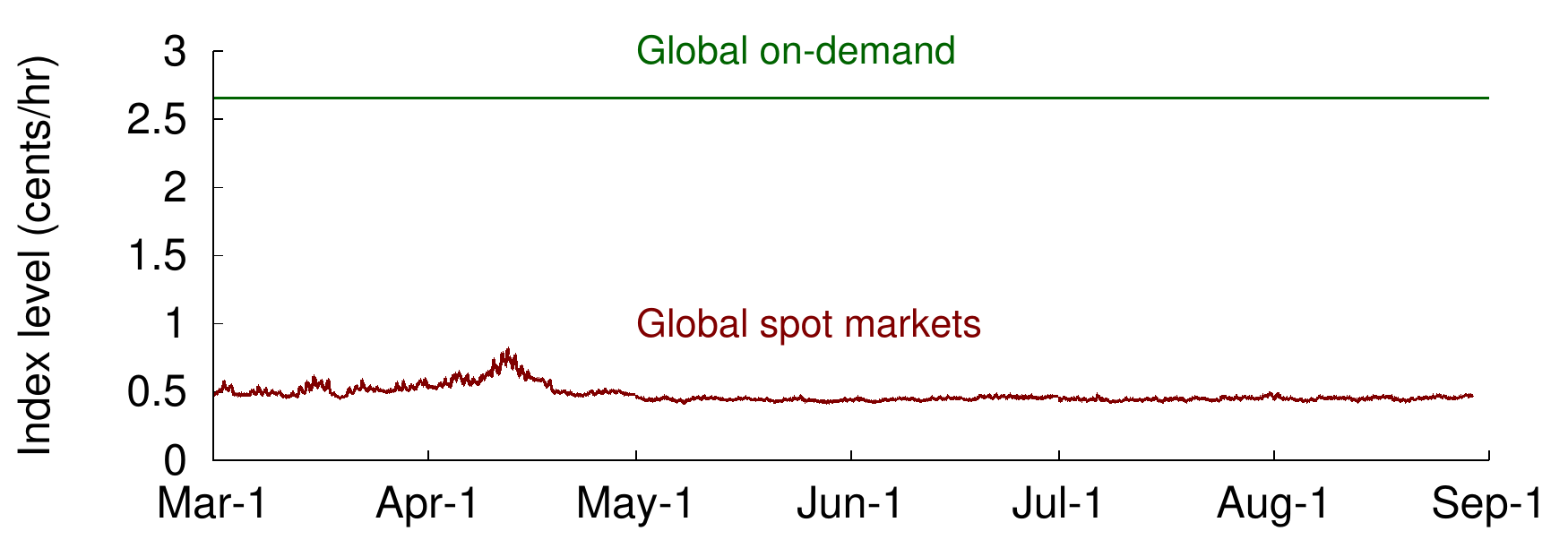} 
\caption{Index price level and spread across all Linux spot VMs in EC2 (2406 over 14 regions).}
\label{fig:global-spot-index}
\end{figure}

Real-world indices must justify their composition, weighting, and consistency.  In the case of a cloud index, these attributes are straightforward to select and justify. The composition determines the set of spot VMs to include in the index.  As discussed previously, two natural groupings would be to include all VMs within a data center (or AZ) or all VMs within a VM family.  The composition could also be tailored to the resource constraints of a particular application.  The weighting defines the relative impact that each spot VM's price has on the index value.  Real-world indices often use either equal weighting, which weights each item equally, or size-proportional weighting, which weights each item by some size metrics, such as volume.  Since EC2 does not publish any details about the volume (or number) of different VMs, we use equal weighting.   Finally, consistency quantifies the stability of the index to changes, i.e., the addition or removal of items, over time. Since we normalize prices and weight spot VMs equally, introducing new spot VMs or removing deprecated ones is possible without altering the intuitive meaning of the index, i.e., in that it still represents the normalized average per-unit cost of spot VM resources.  In the past, EC2 has imposed a cap of 10$\times$ the on-demand price for spot VMs to communicate their temporary unavailability. To ensure consistency, we exclude prices as long as they are equal to this cap.  Given this, we define our cloud index $\mathbb{I}(t)$ below for a selected group of $N$ VMs.  Note that the index price is a relative measure of cost, since its units and absolute value are based on the geometric mean of CPU and memory capacity above. 

\begin{figure}[t]
\centering
\includegraphics[width=0.45\textwidth]{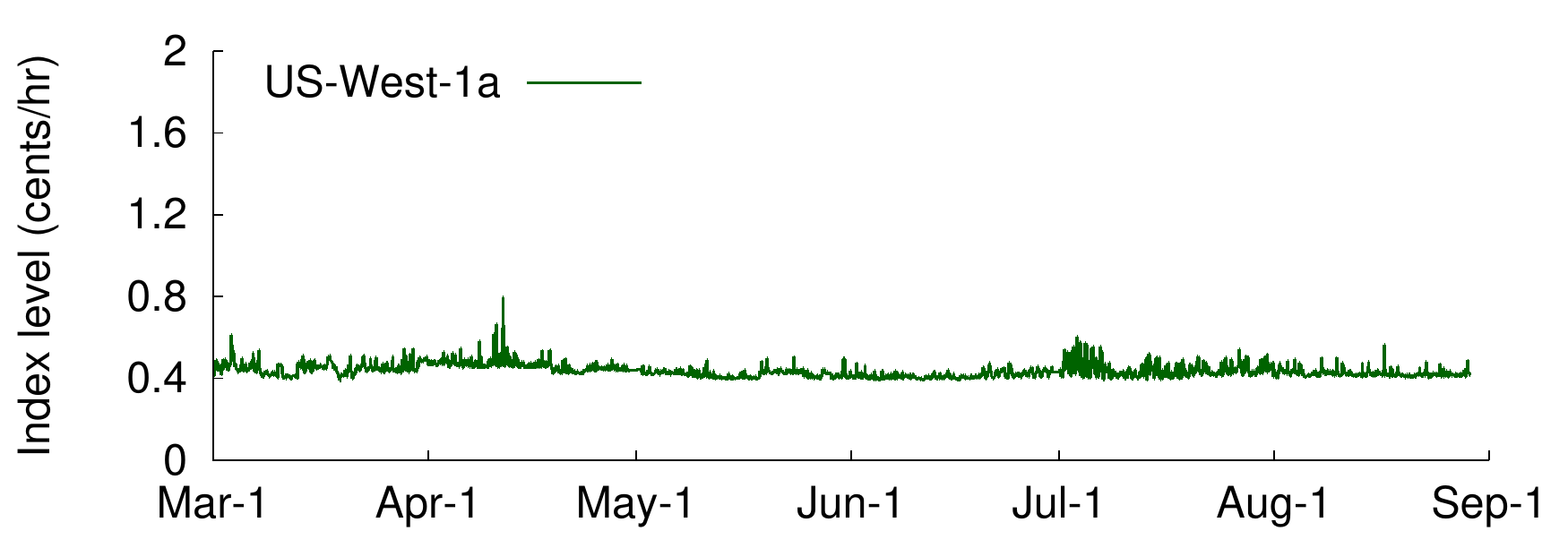} \\
\includegraphics[width=0.45\textwidth]{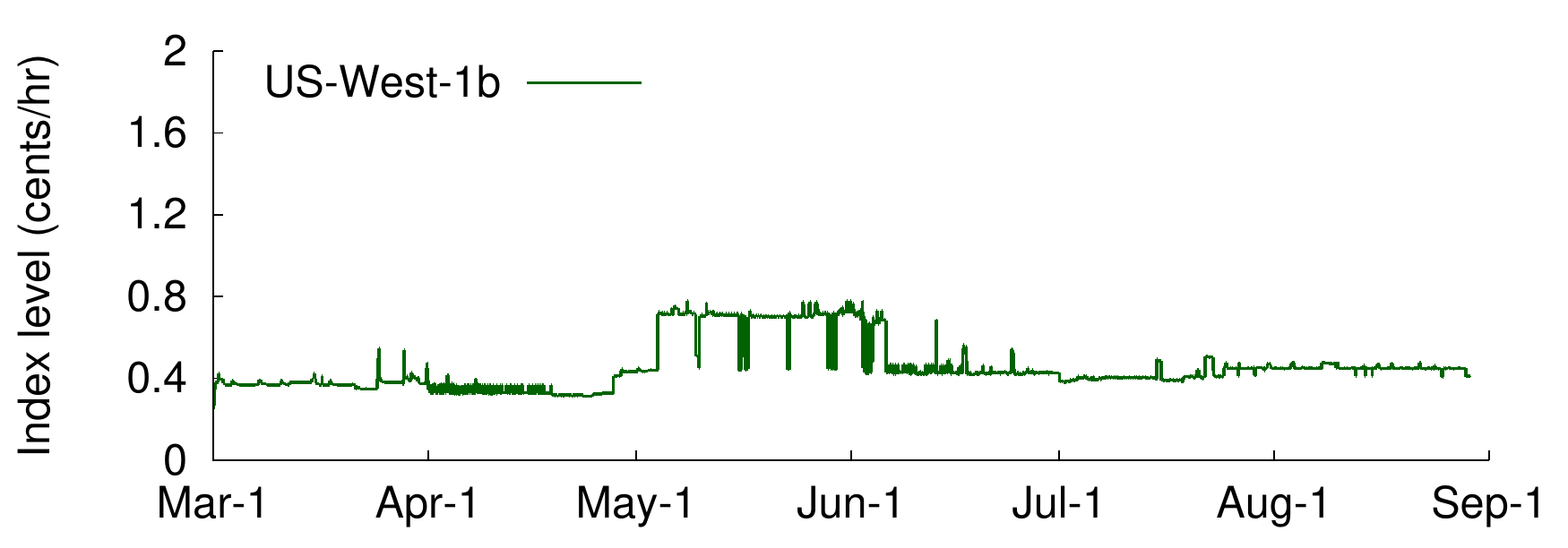} \\
\includegraphics[width=0.45\textwidth]{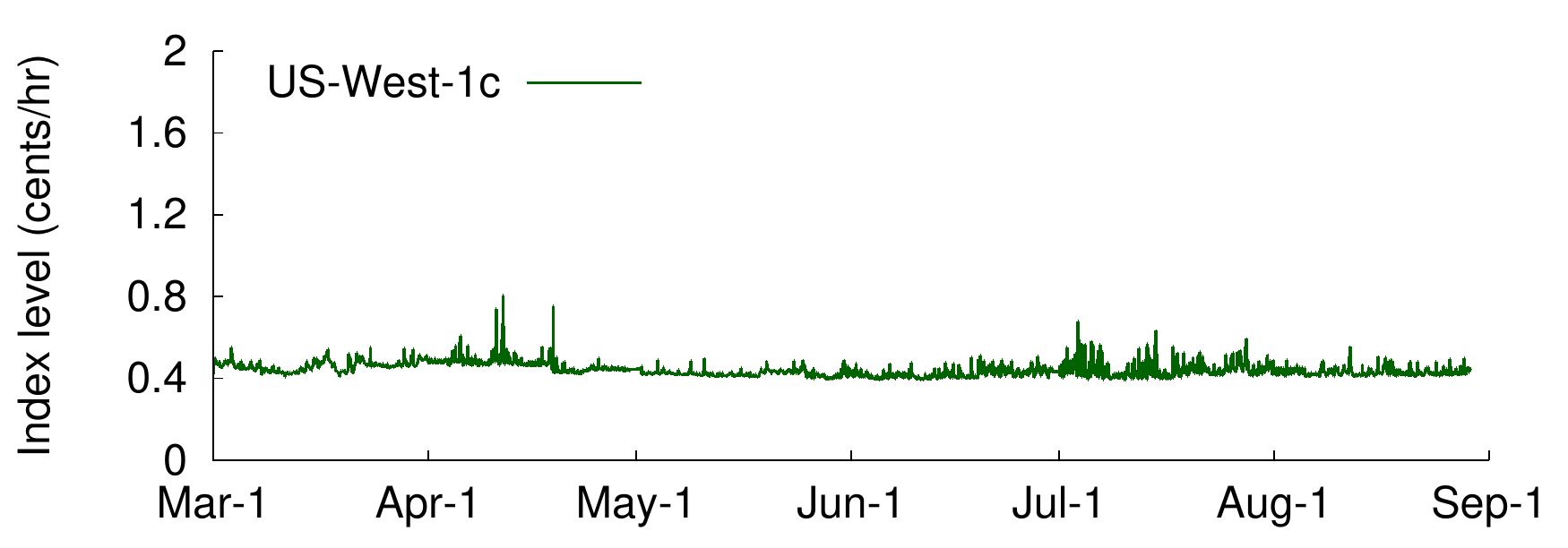}
\caption{Index price for 3 AZs within the {\tt us-west-1} region.}
\label{fig:zone-indices}
\end{figure}

\vspace{-0.3cm}
\begin{equation}
\label{eq:index-definition}
\mathbb{I}(t) = \frac{\sum\limits_{i=1}^{N} {\hat{P}_i(t)}}{N}
\end{equation}

\subsection{Spot Price Data Analysis}

We apply our cloud index above to historical spot prices to quantify their stability and predictability at different levels of aggregation, based on the insights from \Section\ref{sec:first-principles}, to understand how the index can aid in decision making.  We focus on data for Linux VMs in select geographical regions.  Even though other OS configurations and regions have different absolute prices, they are qualitatively similar. 

Figure~\ref{fig:global-spot-index} shows the index price for all 2406 active Linux spot VMs worldwide over a six month period in 2017.  The graph shows that average spot prices per unit of resource are highly stable, with prices roughly $0.5$\textcent/hour, or roughly an 80\% discount over a similar index price for Linux on-demand VMs shown at the top of the graph.  Figure~\ref{fig:zone-indices} then shows the index price for Linux spot VMs across three AZs in the {\tt us-west-1} region.  Similar to Figure~\ref{fig:global-spot-index}, the AZ-level index shows much more stable and predictable behavior than the price of individual spot VMs in Figure~\ref{fig:individual}.  While the average index price over the six month period across the AZs is similar, the index price variations are uncorrelated, likely because each AZ experiences its own local supply and demand dynamics.

\begin{figure}[t]
\centering
\includegraphics[width=0.45\textwidth]{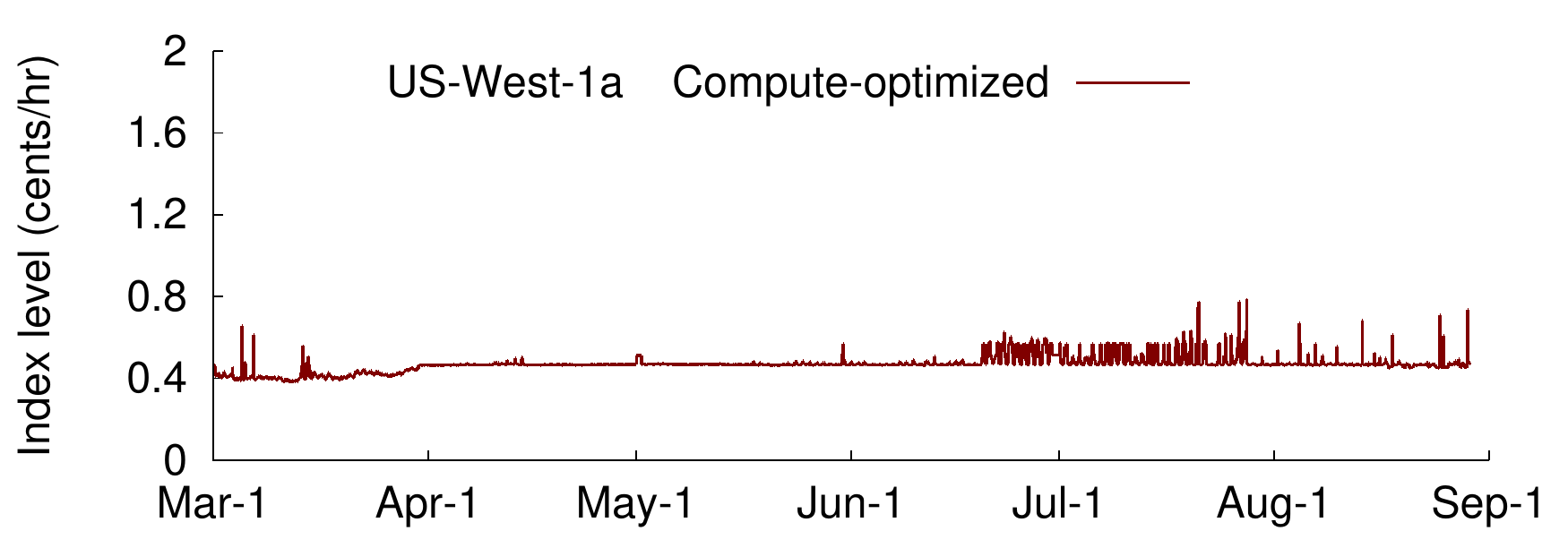} \\
\includegraphics[width=0.45\textwidth]{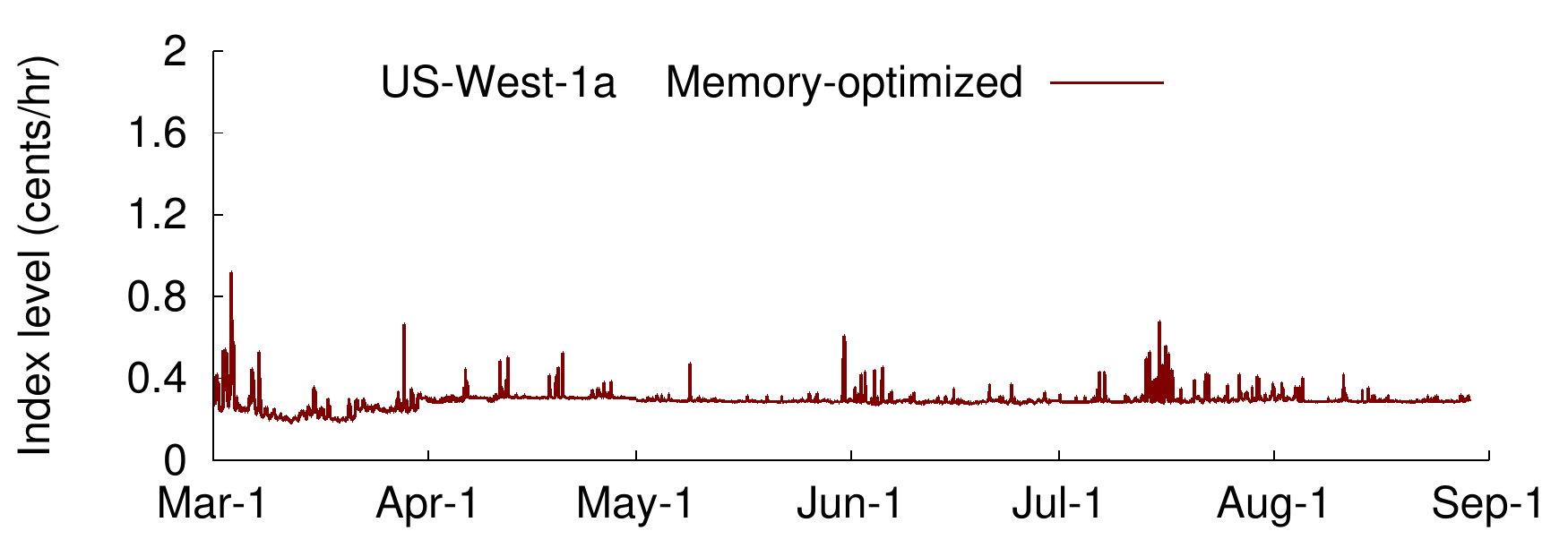} \\
\includegraphics[width=0.45\textwidth]{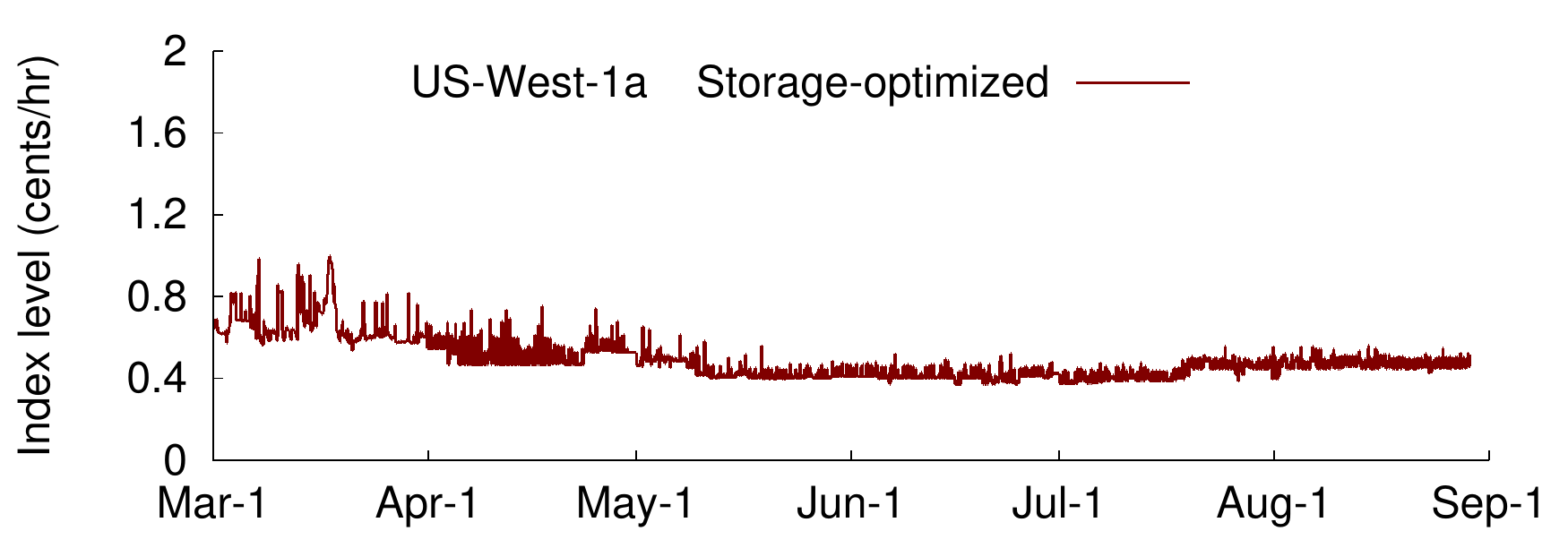}
\caption{Index price for different families of VMs in EC2.}
\label{fig:family-indices}
\end{figure}

\begin{figure}[t]
\centering
\includegraphics[width=0.45\textwidth]{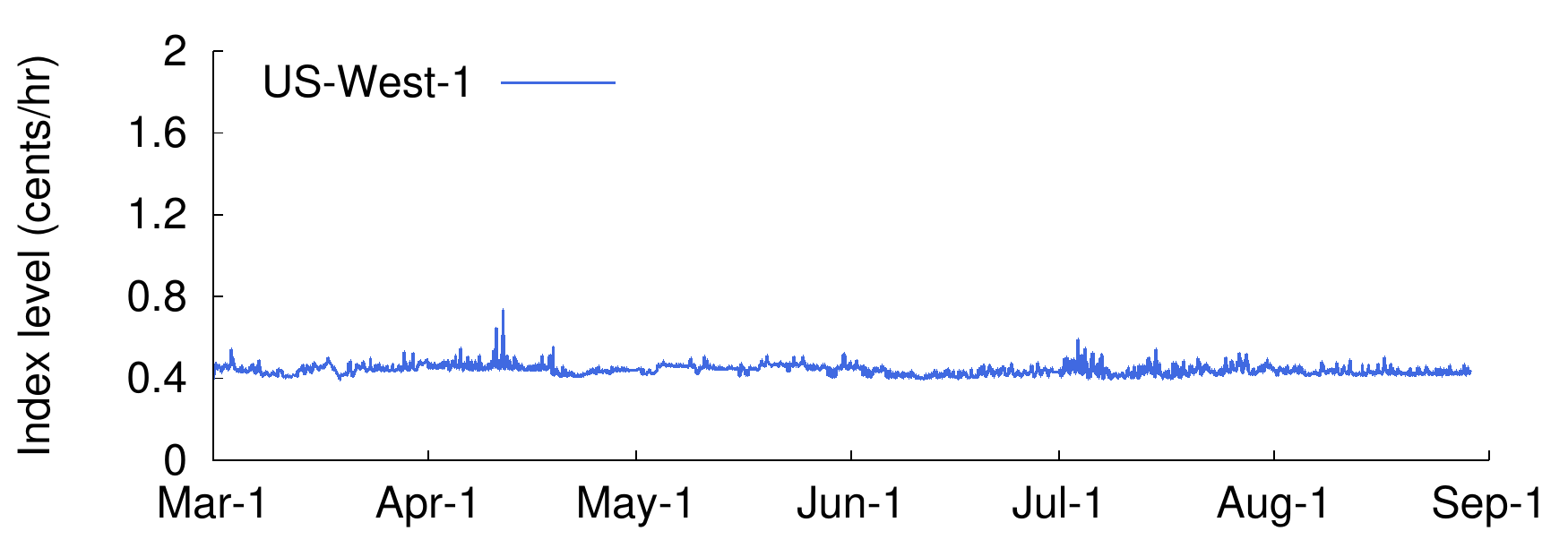}
\caption{Spot index prices at the regional level}
\label{fig:region-indices}
\end{figure}

We plot similar index prices across both VM families and a region.  Figure~\ref{fig:family-indices} shows index prices for compute-, memory-, and storage-optimized VMs in the {\tt us-west-1a} AZ.  While the volatility of the index price for VM families is slightly higher than at the AZ-level, likely because it aggregates over fewer spot VMs, it also remains largely stable and predictable. Similarly, Figure~\ref{fig:region-indices} shows the corresponding regional price index for the {\tt us-west-1} region, which, as expected, shows slightly less volatility than the per-AZ prices in Figure~\ref{fig:zone-indices} due to more aggregation.  In this case, the per-unit regional spot index price is $\sim$16\% less than that of EC2's global spot index price, indicating that {\tt us-west-1} has a lower price than average. 

\noindent {\bf Summary}. \emph{Our data analysis above shows that spot index prices are highly stable and predictable even when the underlying individual spot prices are stochastic. We observe this trend consistently for groups of spot VMs at the global, regional, AZ, and family levels.}

\begin{figure}[t]
\centering
\includegraphics[width=0.45\textwidth]{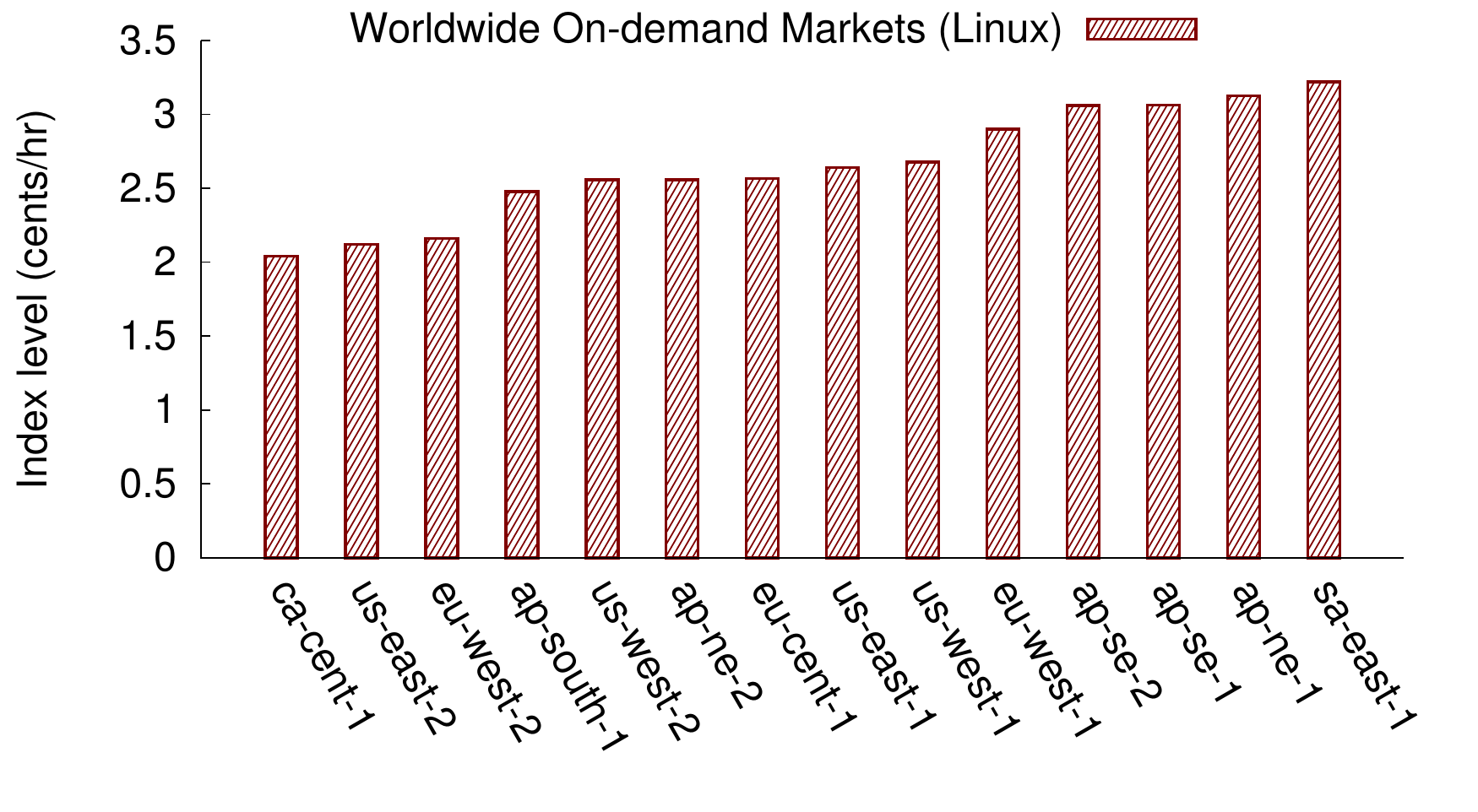} 
\caption{On-demand index prices vary across regions.}
\label{fig:on-demand-index}
\vspace{-0.1cm} 
\end{figure}

As noted above, we can also compute an index price for on-demand VMs in EC2. While EC2's on-demand price for each VM type is fixed within each region, it varies across regions, as shown in Figure~\ref{fig:on-demand-index}, which plots the index price across all 14 regions in EC2.  The figure shows that on-demand index prices vary widely across regions with {\tt sa-east-1} having a 57\% higher price than {\tt ca-central-1}.  Interestingly, significant price differentials exist for regions in close geographic proximity.  For example, the index price for the {\tt us-east-1} region in Virginia is $\sim$20\% higher than that of the {\tt us-east-2} region in Ohio. Such disparities may be due to regional economic factors, such as energy prices, labor costs, or climate.  As with spot VMs, the price differential between on-demand VMs presents a savings opportunity for location-agnostic applications.

\begin{figure}[t]
\centering
\includegraphics[width=0.45\textwidth]{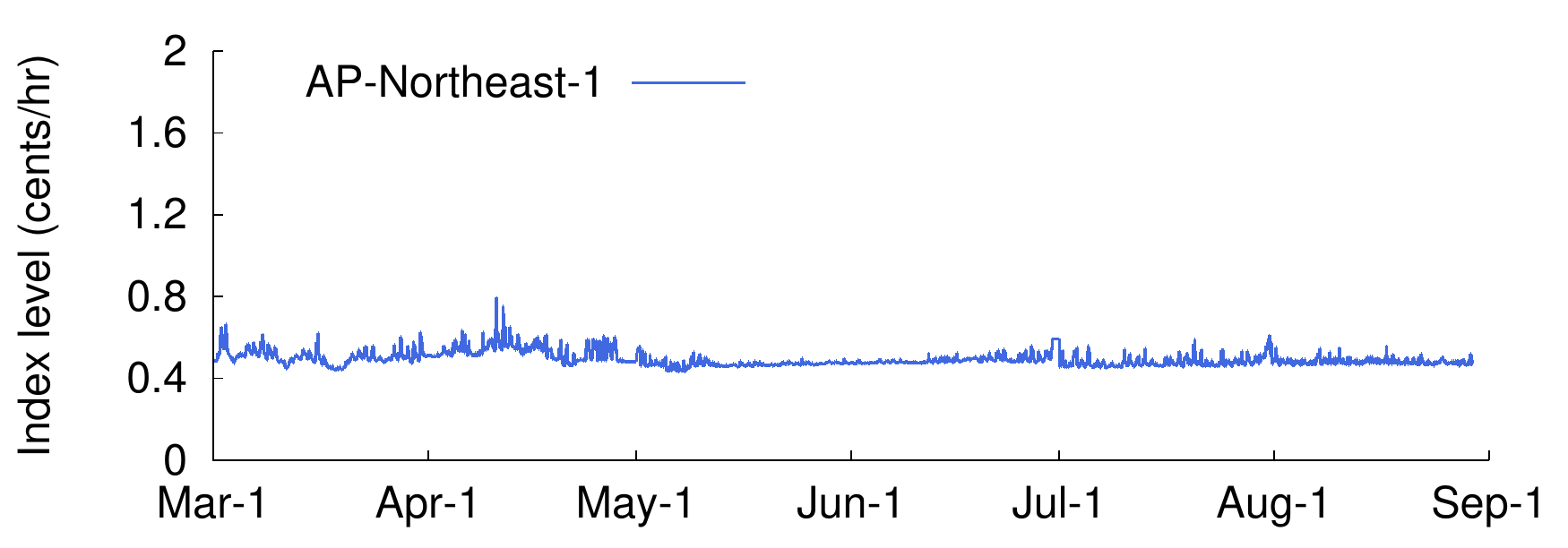} \\
\includegraphics[width=0.45\textwidth]{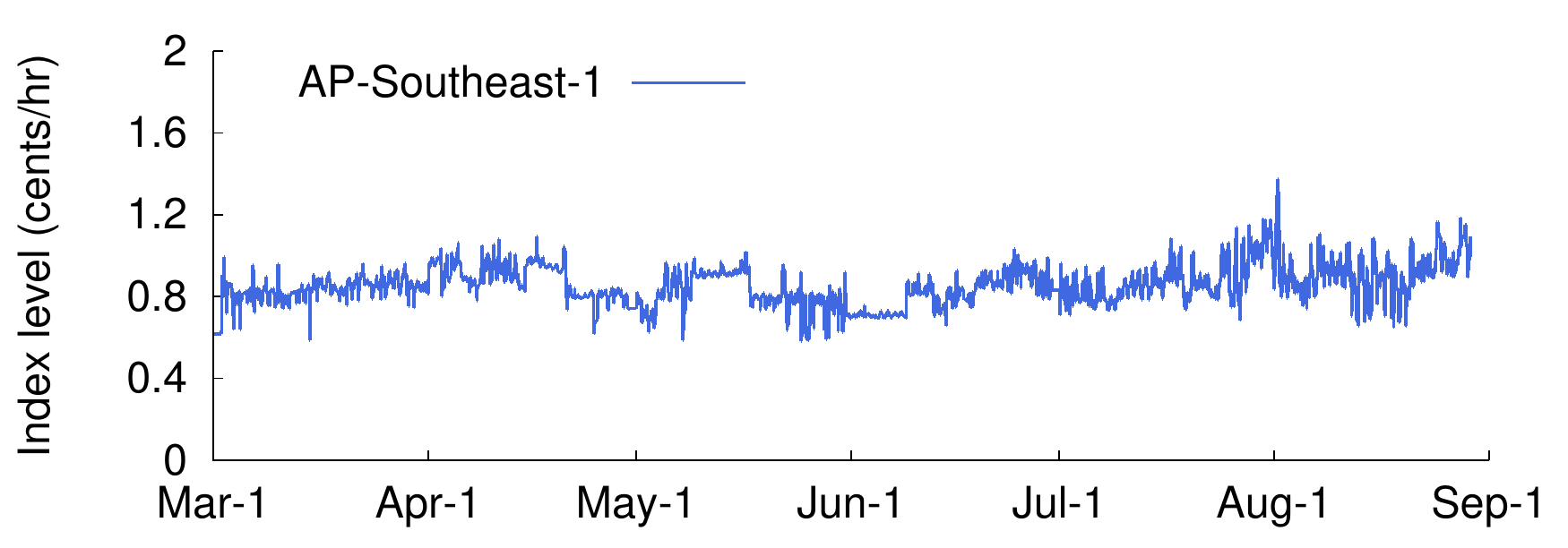}
\caption{Indices showing price variations across regions.}
\label{fig:price-inversion}
\end{figure}

In addition, since on-demand index prices vary across regions, the relative cost savings from using spot VMs in any region also varies.  For example, while the index levels of the {\tt eu-west-1} and {\tt eu-west-2} regions (not shown here) are  $\sim$0.3 and $\sim$0.45\textcent/hour on average, respectively, indicating a 33\% price differential in spot prices, the same price differential is reflected in their on-demand prices.  As a result, the relative percentage savings from using spot VMs over on-demand VMs is the same in each region, although {\tt eu-west-1} has lower absolute prices. Similar regional price inversions also occur between on-demand and spot VMs.  For example, though on-demand VMs in {\tt ap-northeast-1} are slightly more expensive than in {\tt ap-southeast-1}, spot VMs are cheaper, with the {\tt ap-northeast-1} region exhibiting a 60\% discount over the {\tt ap-southeast-1} region, as shown in Figure~\ref{fig:price-inversion}. 

\noindent {\bf Summary}. \emph{Index prices enable users to identify systematic price differentials, price inversions, and arbitrage opportunities between different spot VMs and on-demand VMs across AZs and regions.}

\section{Cloud Index Tracking}
\label{sec:design}

The previous section defined the notion of a cloud index and showed that cloud index prices are more stable and predictable than individual spot prices.   In this section, we leverage this insight to design an approach to cloud index tracking, which runs applications on variable-priced spot VMs such that they incur a predictable cost.  Of course, running embarrassingly parallel, large-scale applications for a predictable cost is trivial, since they can simply select the different spot VMs that comprise the index.  As a result, we focus on smaller scale applications composed of a few VMs that cannot rely on their sheer scale to smooth their costs.  

Our approach to cloud index tracking is simple: we first filter the set of candidate spot VMs that satisfy an application's resource requirements, and then compute a cloud index on the remaining spot VMs, a described in Equation~\ref{eq:index-definition}, to obtain a target cost. We then select the ``best'' spot VM that meets some target objective; \Section\ref{sec:selection-policies} outlines different spot VM selection policies that yield different tradeoffs between VM availability and cost objectives.  If prices or an application's workload changes, such that the selected spot VM no longer satisfies the target objective, then we transparently migrate the application to another spot VM that does satisfy it. As we discuss in \Section\ref{sec:track-hop-property}, such a spot VM should always exist.   Cloud index tracking is reminiscent of both index tracking in finance, which constructs securities to track the price of a reference index, and active trading, which buys and sells stocks over short time horizons to benefit from price volatility.  Of course, there are differences between these financial products and compute resources, as we discuss below. 

\subsection{Index Tracking by Server Hopping}
\label{sec:track-hop-property}

In finance, index tracking is a rule-based investment mechanism that seeks to match the financial returns from a portfolio to the performance of the price index that it tracks. Index tracking enables investors to ``buy the market,'' rather than attempting to ``beat the market'' by selecting their own individual investments.  Index tracking has become increasingly popular, since if prices fully reflect all available information, as assumed by the efficient-market hypothesis~\cite{efficient-market-hypothesis}, then consistently beating the market by accurately predicting prices is impossible.  Our approach derives from the same intuition: while it is difficult to predict the price of individual spot VMs, the index price of groups of spot VMs are more predictable.  However, our approach focuses on matching a single VM's cost to the index price, rather than a large group of VMs.  To track the cost benefit of a selected spot VM $i$ with respect to a reference cloud index $\mathbb{I}$, we define the equation below, where $\mathbb{G}ain(t_1,t_2)$ represents the gain in the index between times $t_1$ and $t_2$ that spot VM $i$ was held, while $\hat{P}_i$, $C_i$ and $M_i$ denote the spot VM's normalized index price, CPU, and memory capacity, respectively.

\vspace{-0.3cm}
\begin{equation}
\label{eq:gain-on-index}
\mathbb{G}ain(t_1,t_2) = \sum\limits_{t=t_1}^{t_2} (\mathbb{I}(t) - \hat{P}_i(t)) \cdot \sqrt{C_i \cdot M_i} 
\end{equation}

The equation above simply denormalizes the index price $\mathbb{I}$ in Equation~\ref{eq:index-definition} to compute the gain in cost relative to it for a VM with a specific CPU and memory capacity.  That is, if the normalized price of spot VM $i$ is less than that of the cloud index, then the gain is positive. Note that in the equation $\hat{P}_i(t)$ essentially represents an index price for only spot VM $i$.  Of course, since a single spot VM's price may change over time, maintaining a positive gain relative to the index price may require migrating to another server.

Recent advances in container and nested virtualization, datacenter networking, and fine-grained, i.e., per-second, billing models enable applications to frequently migrate from one cloud VM to another in response to real-time price and workload dynamics.  For example, prior work on Superclouds~\cite{supercloud} live migrates applications in response to geographically shifting workloads, while HotSpot~\cite{hotspot} migrates applications to the most cost-efficient spot VM.  However, such dynamic migration policies are designed as localized greedy optimizations: they incur upfront migration costs, e.g., downtime or degraded performance, for a future benefit, e.g., better performance or a lower cost.  In contrast, our primary goal is ensuring predictable costs by\ preventing a spot VM's cost from significantly deviating from the index cost.  In this case, every migration reduces the accrued gain relative to the index from Equation~\ref{eq:gain-on-index}.  To account for this loss, we compute the overhead of migration, which requires paying for two spot VMs $i$ and $j$ for the migration's duration $T_m$, and accounting for the lost work over that time. Thus, we compute $\mathbb{L}oss(i,j)$ from migrating from $i$ to $j$ as below. 

\vspace{-0.3cm}
\begin{equation}
\label{eq:loss-on-hopping}
\mathbb{L}oss(i,j) = ({P}_i(t) + {P}_j(t)) \cdot {T}_m
\end{equation}

By tracking an application's $\mathbb{G}ain$ and $\mathbb{L}oss$ over time using Equation's \ref{eq:gain-on-index} and \ref{eq:loss-on-hopping}, we can compute its cost relative to the index.   If the cost relative to the index is negative, then we should migrate the application to maintain the index price.  Since the index price is simply an average across many spot VMs, there must be spot VMs with a price equal to or below the index price.  As a result, if the price of the current spot VM rises, there should always be a cost-efficient option for migration.  The equation computes the monetary cost of the migration, i.e., the cost to maintain two VMs over its duration.  

\begin{figure}[t]
\centering
\includegraphics[width=0.45\textwidth]{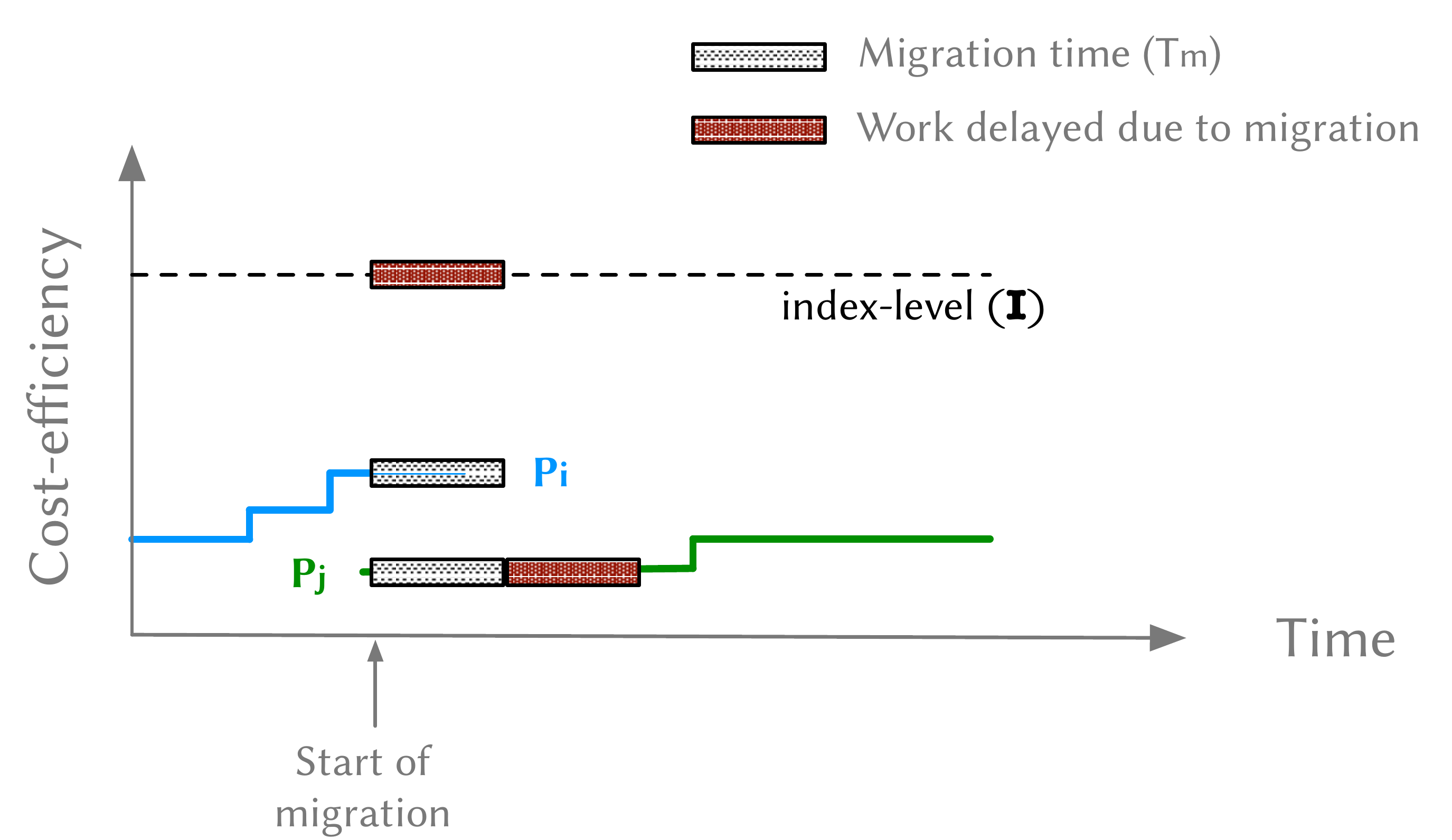}
\caption{Illustration of the overheads of migration, including double-paying for VMs and lost work.}
\label{fig:sufficiency-condition}
\end{figure}

To ensure that we maintain the index price when migrating, we only trigger migrations when the total cost of the migration is less than the index price, as shown in Equation~\ref{eq:sufficient-condition} below.  The term on the right sums the cost of source and destination VMs during the migration, as well as the cost to execute the work delayed by the migration (resulting in the 2 factor).  Assuming a simple stop-and-copy migration, as generally implemented by resource containers, the application does no work over the duration of the migration.  In addition, as we discuss next, applications are also not available during the stop-and-copy migration. Figure~\ref{fig:sufficiency-condition} illustrates this loss of work and unavailability when migrating from VM $i$ to $j$ over time $T_m$ (shown by the gray area).   These losses can decrease the accrued $\mathbb{G}ain$s from above under aggressive migration policies that chase low prices by migrating to the lowest cost VM.  Migrating under the condition below ensures that migrations only occur when they maintain the index level.  The equation simply accounts for the cost of the migration and the lost work above, and then re-normalizes the prices for comparison with the index price. 

\vspace{-0.3cm}
\begin{equation}
\label{eq:sufficient-condition}
\mathbb{I}(t) > \hat{P}_i(t) + (2 \cdot \hat{P}_j(t)) 
\end{equation}

One advantage from cloud index tracking relative to greedy approaches is that it should scale well with increased adoption.  If everyone adopts greedy approaches that chase low prices by aggressively migrating to the lowest cost VM, then prices are likely to become more volatile.  In contrast, if everyone adopts index tracking, spot VM prices should become increasingly stable as everyone's target price represents the fair market value of the idle cloud capacity. This stability should reduce the migrations required to maintain the index price, resulting in less lost work and higher availability. 

\begin{figure*}[t]
\centering
\includegraphics[width=0.8\textwidth]{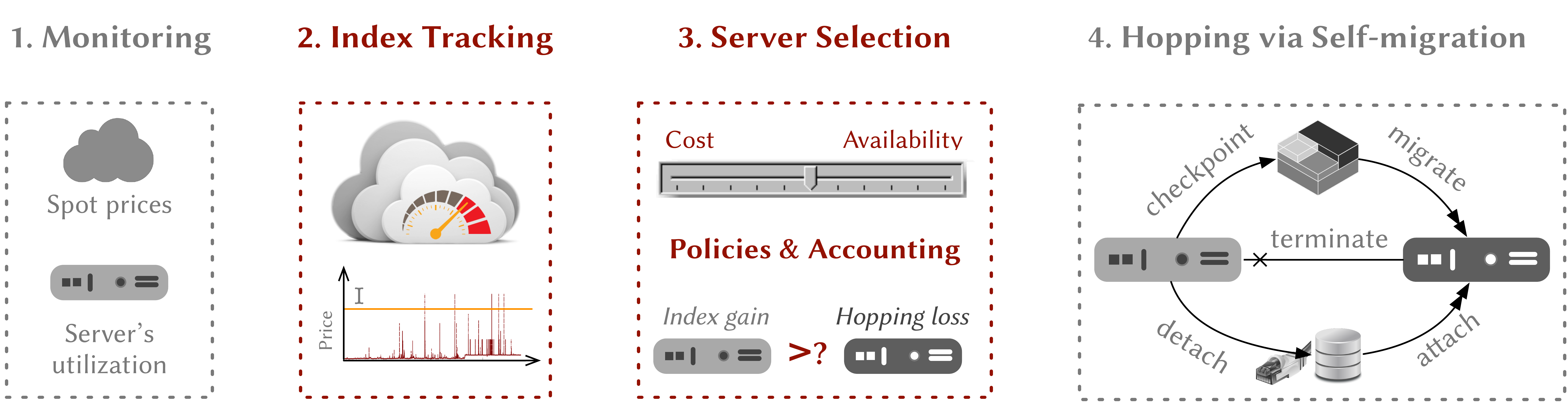}
\caption{System architecture with HotSpot components boxed in {\color{gray}gray} and our extensions for cloud index tracking in {\color{Maroon}red}.}
\label{fig:hotspot-extensions}
\vspace{-0.2cm}
\end{figure*}

\subsection{Selection and Migration Policies}
\label{sec:selection-policies}

Before presenting our cloud index tracking policy below, we first present two other migration policies that expose a tradeoff between cost and availability.  We compute the cost of using the policies based on their gain and loss using the equations in the previous section.  We compute availability based on the number of  migrations and their associated downtime, as depicted in Figure~\ref{fig:sufficiency-condition}.

\noindent{\bf Cost-centric Policy.}  This policy's goal is to greedily maximize cost savings by aggressively migrating to the spot VM that is most cost-efficient when accounting for the application's resource utilization, as described in prior work~\cite{hotspot}.   To account for resource utilization, we re-normalize the VM's cost from Equation~\ref{eq:cost-efficiency} based on the actual resources utilized at time $t$, namely $C_{utilized}$ and $M_{utilized}$. 

\begin{equation}
\label{eq:cost-eff-utilized}
\breve{P}_i(t) = \frac{P_i(t)}{\sqrt{C_{utilized} \cdot M_{utilized}}}
\end{equation}
\vspace{0.1cm}

The cost-centric policy then chooses the spot VM $i$ that provides the best value $\breve{P}_i$ at time $t$.  The policy only optimizes for cost, and does not consider availability.  The policy selects and migrates to a new spot VM anytime it observes a lower cost spot VM emerge due to changes in spot prices or an application's resource utilization, and the cost overhead of migration is less than the expected cost benefit from the lower price.   Note that the cost-centric policy does not consider the index price or Equation~\ref{fig:sufficiency-condition} when triggering a new selection and migration. 

\noindent{\bf Availability-aware Policy.}  This policy's goal is to maximize availability, rather than cost savings, by minimizing the number of migrations, and the associated downtime.  To do so, the policy seeks out spot VMs with the most stable prices that also yield a cost less than the index price. To identify such spot VMs, the policy computes the standard deviation of each spot VM's price over a pre-defined window in the past.  Then, from among the spot VMs with a price below the index price, it picks the one with the lowest standard deviation.   As above, the policy computes the price with respect to the resources the application is utilizing at time $t$.  As noted above, there will always be at least one spot VM with a price below the index price.  The policy does not trigger a new selection and migration until its current spot VM's price rises above the index price.  As a result, the policy only migrates when necessary to maintain the index price, and primarily selects VM's based on their price stability and thus availability. 

\noindent{\bf Balanced Index Tracking Policy.} We define our cloud index tracking policy to be balanced and mind the gap between the two policy extremes above by considering both cost and availability when making decisions.  To do so, we observe that the higher the variability in a VM's spot price, the higher the probability of triggering a migration to another VM to maintain the index price, which decreases availability.  Thus, variability and availability are related to each other.  In finance, the Sharpe ratio~\cite{sharpe-ratio} is common metric for balancing cost and variability in prices.  The Sharpe ratio $\mathbb{S}_i$ estimates an asset's risk-adjusted returns: for an asset $i$, it is the ratio of the expected difference between the asset's returns $R_i$ and the risk-free returns $R_{free}$ divided by the standard deviation of the asset's returns $\sigma_i$.  Here, we adapt the standard Sharpe ratio by replacing $R_{free}$ with our cloud index ($\mathbb{I}$) to represent the returns we should expect, and replacing $R_i$ with $\breve{P}_i$ (or the price of the current VM per unit of resource utilized), where $\sigma_i$ is the standard deviation of $\breve{P}_i$.  We compute $\breve{P}_i$ and $sigma_i$ over the same window. 

\vspace{-0.1cm}
\begin{equation}
\mathbb{S}_i(t) = \frac{\mathbb{I}_{g}(t) - \breve{P}_i(t)}{\sigma_i}
\end{equation}
\vspace{0.1cm}

The numerator estimates a VM's ``return'' relative to the index price, and the denominator quantifies its ``risk'' of deviating from the index price and thus requiring a migration that results in unavailability.  Higher values of this ratio are better, since they indicate a low cost and a high stability, i.e., availability.  Thus, this policy triggers a migration when the current VM no longer has the highest modified Sharpe ratio, or balance factor, and Equation~\ref{eq:sufficient-condition} holds. 

\section{Implementation}
\label{sec:implementation}

Our implementation builds on prior work on dynamic cost-driven migration in HotSpot, which implements the reactive cost-centric migration policy described in the previous section~\cite{hotspot}.  HotSpot applications run in containers that are self-migrating: a daemon runs on the VM (external to the container) and operates a feedback loop that continuously monitors and analyzes spot prices and application resource utilization, and aggressively migrates to the lowest cost spot VM.  HotSpot already includes the mechanics for price and utilization monitoring, requesting a new spot VM and terminating a previous one, and migrating a network-accessible container between spot VMs (using a stop-and-copy migration).  The HotSpot daemon is embedded into an Amazon Machine Image (AMI) external to the container, and starts on boot-up. As a result, the system requires no external infrastructure, e.g., such as a remote master server that monitors prices and triggers migrations. 

We build on HotSpot to implement cloud index tracking in this paper. Specifically, our implementation runs applications inside of Linux Containers (LXC), which only reliably support stop-and-copy migration and not live migration. In addition, migrations require applications to use remote storage, i.e., Amazon's Elastic Block Store (EBS), and virtual networking, i.e., Amazon's Elastic Network Interface (ENI).  Each containerized application manages itself without any external coordination with applications running on remote VMs.  Hotspot was implemented in {\tt python} and integrates with EC2's {\tt boto3} library, LXC bindings, and various administrative shell scripts.  We retain the monitoring and migration components but replace HotSpot's greedy cost-centric migration policy with our cloud index tracking policy from the previous section.

\begin{figure*}[t]
\makebox[1\textwidth][c]{
\centering
\begin{minipage}[c]{1\textwidth}
\hspace{0.5cm}
\includegraphics[width=0.4\textwidth]{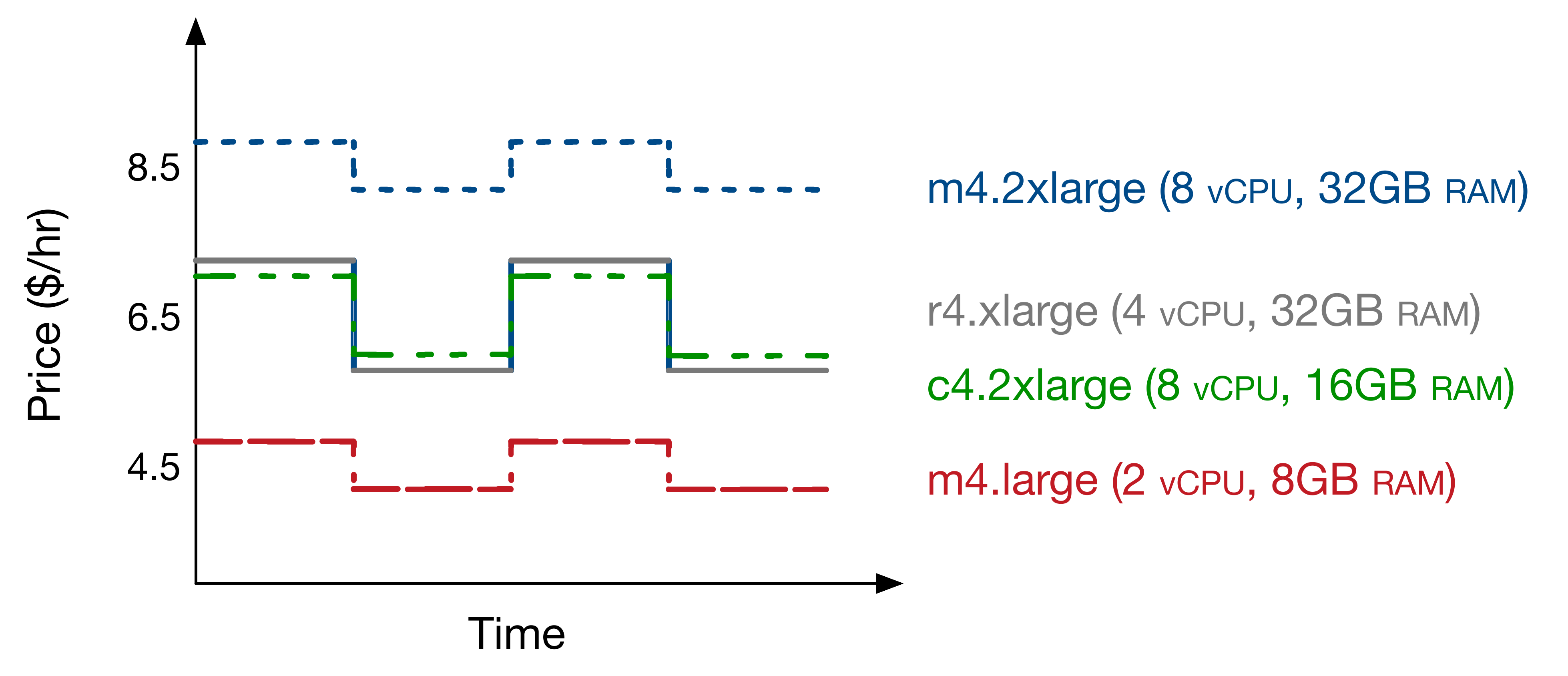}
\hspace{1cm}
\includegraphics[width=0.2\textwidth]{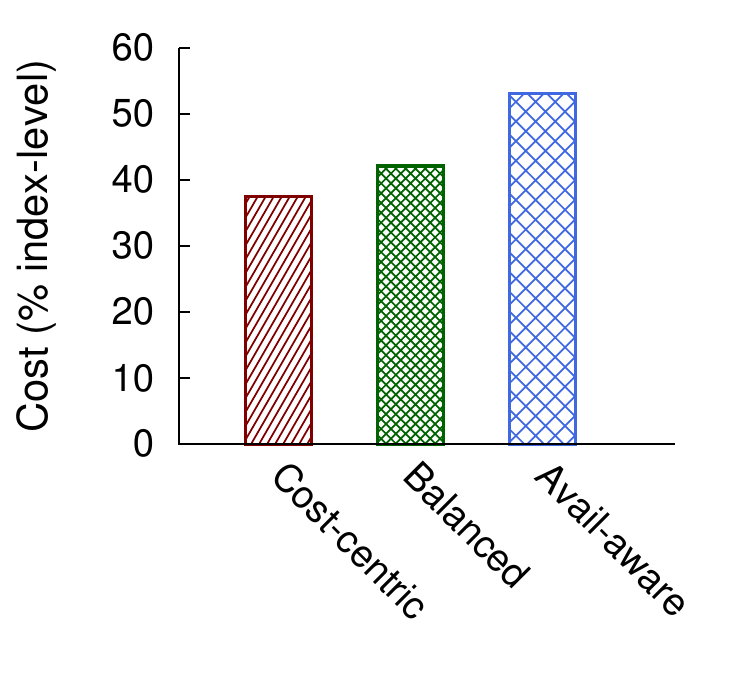}
\includegraphics[width=0.2\textwidth]{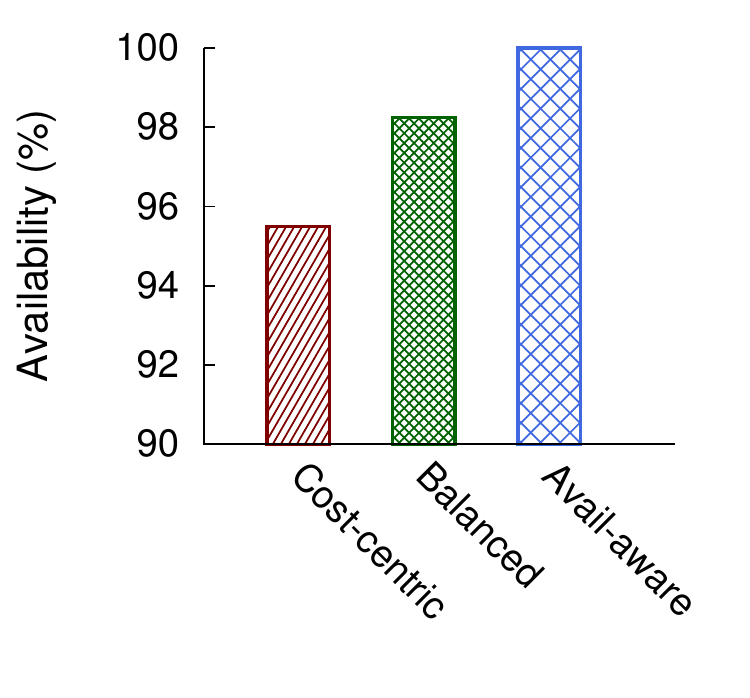}
\end{minipage}}
\caption{Synthetic spot prices (left), and the cost-availability tradeoffs of different policies for our baseline job (right).}
\label{fig:policies-baseline}
\end{figure*}

\begin{figure*}[t]
\centering
\includegraphics[width=0.33\textwidth]{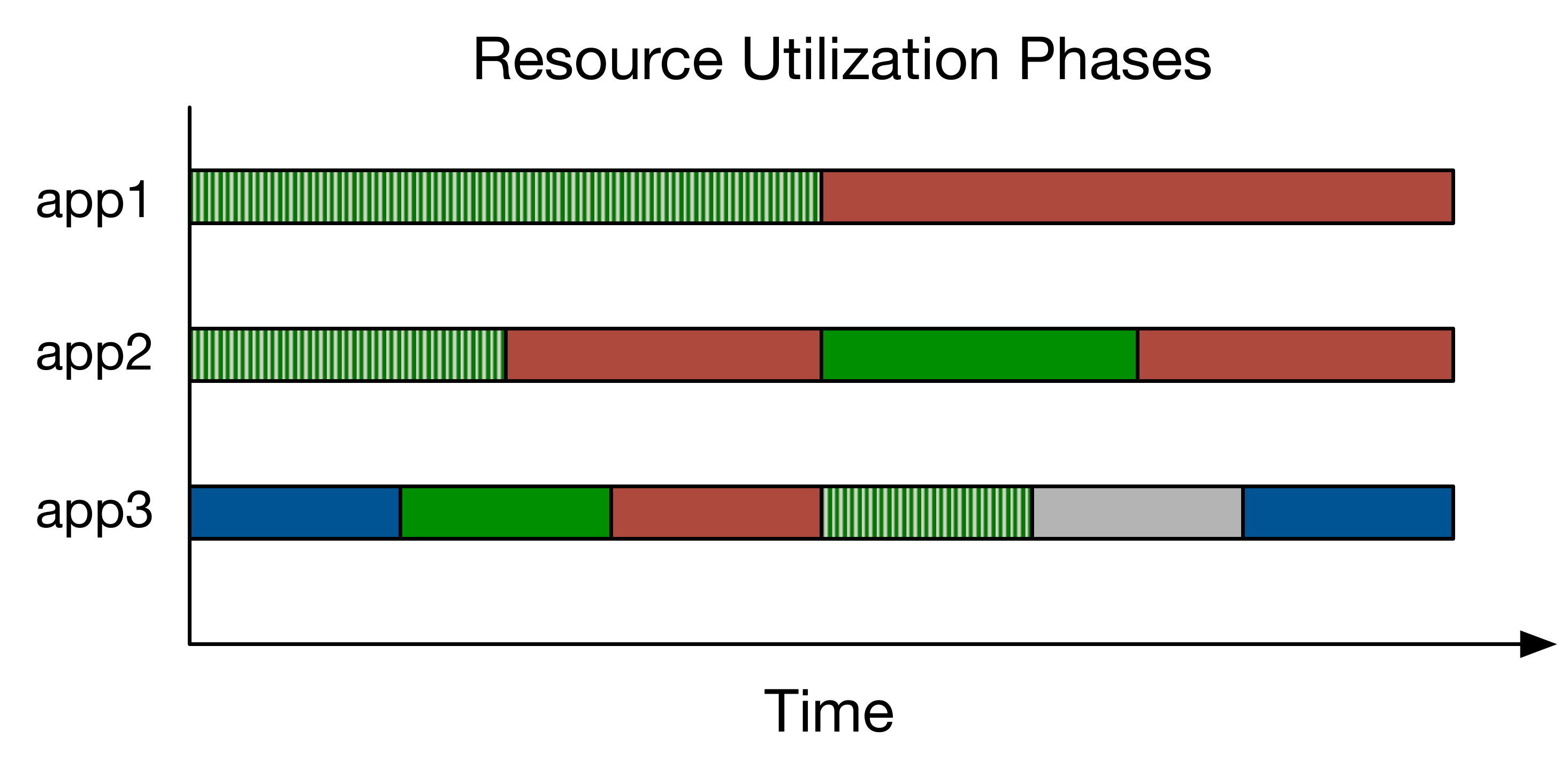}
\hspace{1cm}
\includegraphics[width=0.26\textwidth]{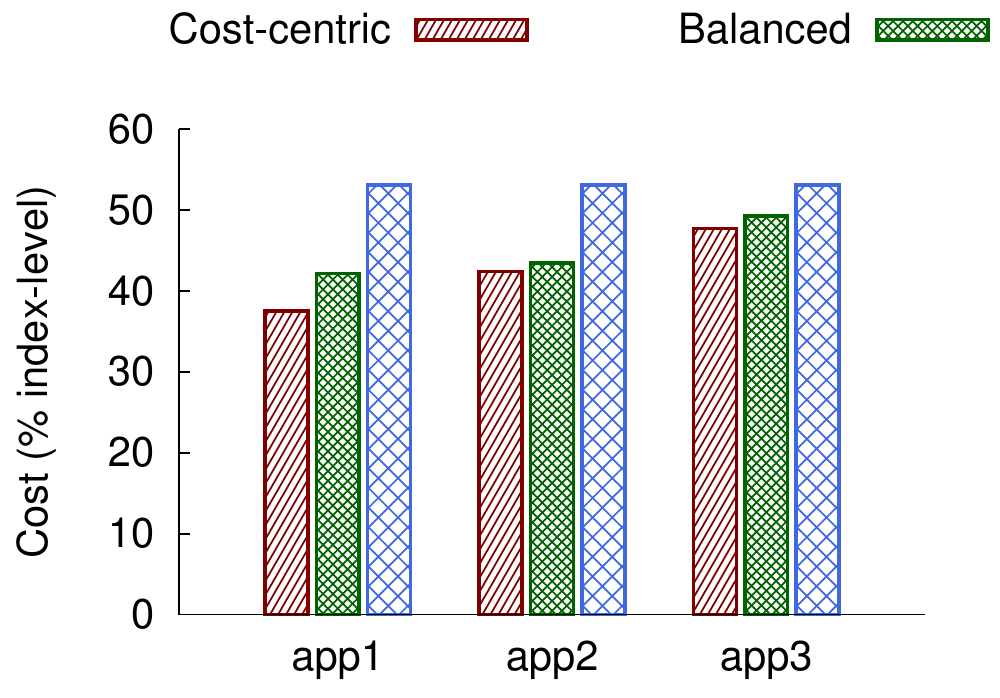}
\includegraphics[width=0.26\textwidth]{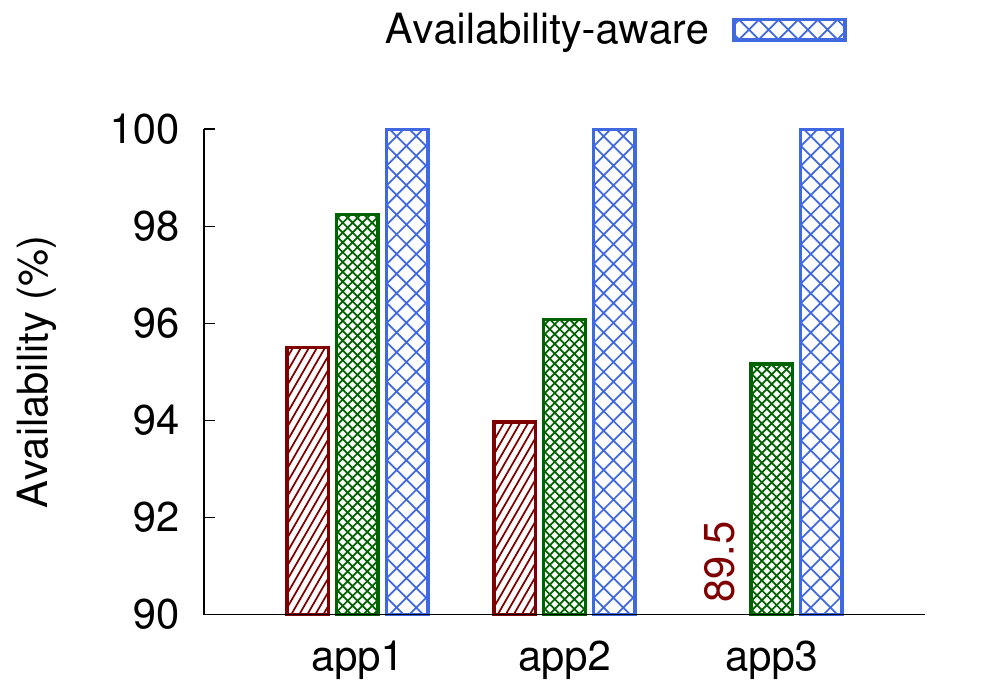}
\caption{Utilization phases for 3 jobs depicted in the left graph. The color of a phase indicates the best matching server type (from Figure~\ref{fig:policies-baseline}), and  green-gray stripes indicate the existence of two best matching servers. The right two graphs show the cost-availability tradeoff for different migration policies for each job.}
\label{fig:policies-app-usage}
\end{figure*}

Figure~\ref{fig:hotspot-extensions} illustrates the modifications and extensions relative to HotSpot. In particular, we implemented a standalone utility that takes as input an application's resource requirements and computes the index price for a selected AZ, which can be used to estimate an application's total cost for a given running time.  Thus, the utility enables users to know their expected cost before starting an application. We integrate a library version of this standalone utility into our cloud index tracking module, which polls the price and utilization monitoring engine once every 5 minutes to update the gain from index tracking. This update also triggers the selection and migration policy, which iterates through all spot VMs that satisfy the resource requirements to determine whether to migrate to a new VM based on the migration policies described in \Section\ref{sec:selection-policies}.  For the most populous AZ ({\tt us-east-1a} with 106 spot VMs), one iteration of this feedback loop, including monitoring, tracking, and server selection, takes an average of $\sim$2s to complete.  The container migration uses a direct memory-to-memory transfer, which saves the container's state to the source VM's memory, transfers it directly to the memory of the destination VM, and then restores it, to minimize the downtime and unavailability from migration.  Migration latencies are $\sim$1s/GB with a migration of a 32GB RAM container taking $\sim$30s, and increasing linearly for larger memory sizes~\cite{hotspot}.

\section{Evaluation}
\label{sec:evaluation}

Our evaluation demonstrates the tradeoff between availability and cost exposed by the migration policies in \Section\ref{sec:selection-policies}, and shows that cloud index tracking's balanced policy combines predictable and low costs with high availability.  We first evaluate cost and availability for all policies under different synthetic price characteristics and application workload scenarios using a real prototype running on EC2. We then use real price data from EC2 and workload traces from a production cluster to drive simulations, and compare cloud index tracking with existing static approaches, such as SpotFleet~\cite{ec2-spot-fleet}, that do not migrate as conditions change, and highly reactive approaches, such as HotSpot~\cite{hotspot}, that continuously chase low prices. 

\subsection{Cost-Availability Tradeoff}

We compare the cost-availability tradeoff of selection and migration policies from \Section\ref{sec:selection-policies} using our prototype implementation on EC2.   We generate synthetic application workloads and pricing signals to enable precise control over the different parameters that affect the cost-availability tradeoff.  As a result, we replace a small set of EC2 API function calls in our prototype, such as those that query real-time spot prices, with simulated calls.  We also run a synthetic job that enables us to precisely control its CPU and memory usage. Specifically, we use the lookbusy synthetic load generatior, which can be configured to consume a precise amount of CPU and memory resources~\cite{lookbusy}.  Our baseline lookbusy job runs for 1 hour on a reference {\tt m4.2xlarge} and has 2 distinct resource utilization phases, each thirty minutes long.  The first phase consumes 4 vCPUs at 100\% utilization and 16GB memory, while the second phase consumes 2 vCPUs and 8GB memory. 

To enable repeatable experiments, we generate synthetic spot price traces for 4 spot VMs: the {\tt m4.large}, {\tt m4.2xlarge}, {\tt c4.2xlarge} and {\tt r4.xlarge}. We choose these VMs, since the vCPU allotment varies between 2 and 8, and their memory capacity varies between 8GB and 32GB, which covers the entire spectrum of our application's resource utilization.  We model their dynamic spot prices as follows: the {\tt m4.large} has an average price of 4.5\textcent/hour with a standard deviation of 0.5; the {\tt m4.2xlarge} has an average price of 8.5\textcent/hour and a standard deviation of 0.5; the  {\tt c4.2xlarge} has an average price of 6.5\textcent/hour with a standard deviation of 1; and the {\tt r4.xlarge} also has an average price of 6.5\textcent/hour with a standard deviation of 1.1. Figure~\ref{fig:policies-baseline}(left) illustrates these spot prices characteristics for each VM.  Note that these average prices are absolute, and cover a range of costs per unit of CPU and memory.  For example, the {\tt c4.2xlarge} is 2\textcent\ cheaper per hour than the {\tt m4.2xlarge}, but has the same vCPU allotment and 16GB less memory. In our experiments, we generate a new spot price uniformly randomly each minute such that the hour-long price trace adheres to the average prices and standard deviations above across all 60 price changes.  As in EC2, we assume a per-second billing model. 

\noindent {\bf Baseline Job Performance}. Figure~\ref{fig:policies-baseline}(right) shows both the cost and availability from running our baseline job above under the 3 different policies---cost-centric, availability-aware, and balanced---from \Section\ref{sec:selection-policies}.  We normalize the cost on the y-axis to that of the reference VM that costs the index-level, i.e., the average across all VMs.  We observe that all policies yield a lower cost than the cloud index, indicating that the price differentials in our synthetic traces are large enough to incentivize even the availability-aware policy, which prioritizes stability, to migrate.  Also, as expected, the cost-centric policy, which aggressively chases low prices, has the lowest overall cost, while the availability-aware policy, which discourages migrations, has the highest availability.  Our balanced policy, which uses a modified Sharpe ratio to consider both cost and price variability, achieves a balance between cost and availability, coming within 12\% of the cost-centric policy's cost and within 1.7\% of the availability-aware policy's availability. 

\noindent {\bf Changing Job Characteristics}. We next modify the baseline job from above to vary its resource utilization.  We depict the modified resource utilization patterns in Figure~\ref{fig:policies-app-usage}(left).  The resource variations are such that the application could be executed on at least one of the four target spot markets at all loads.  Figure~\ref{fig:policies-app-usage}(right) shows the effect on cost and availability from running the three different jobs.  As the figure shows, the availability-aware policy does not change across the jobs, as it optimizes for stability and not cost savings, so it performs no migrations in each case.  However, both the cost-centric policy and the balanced policy incur increasing cost overheads as the application's resource utilization becomes more volatile.  The increased volatility in utilization causes the policies to trigger more migrations, since, even though the spot prices remain the same, the price per unit of utilized resource has become more volatile.  Thus, the cost overhead and downtime due to migrations increases, as the job's resource utilization becomes more volatile.  Since the balanced policy does not react to changes in prices or resource utilization as quickly as the cost-centric policy, as it also considers price volatility in determining whether to migrate, its cost increases are less.  For the same reason, the balanced policy maintains a higher availability compared to the cost-centric policy. 

\begin{figure}[t]
\centering
\includegraphics[width=0.23\textwidth]{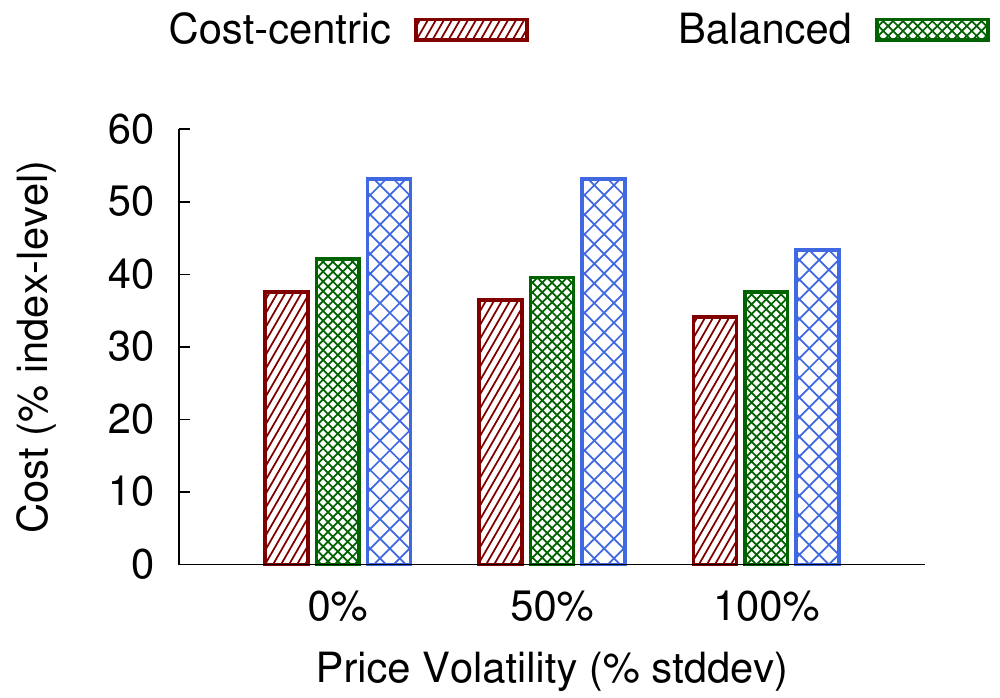}
\includegraphics[width=0.23\textwidth]{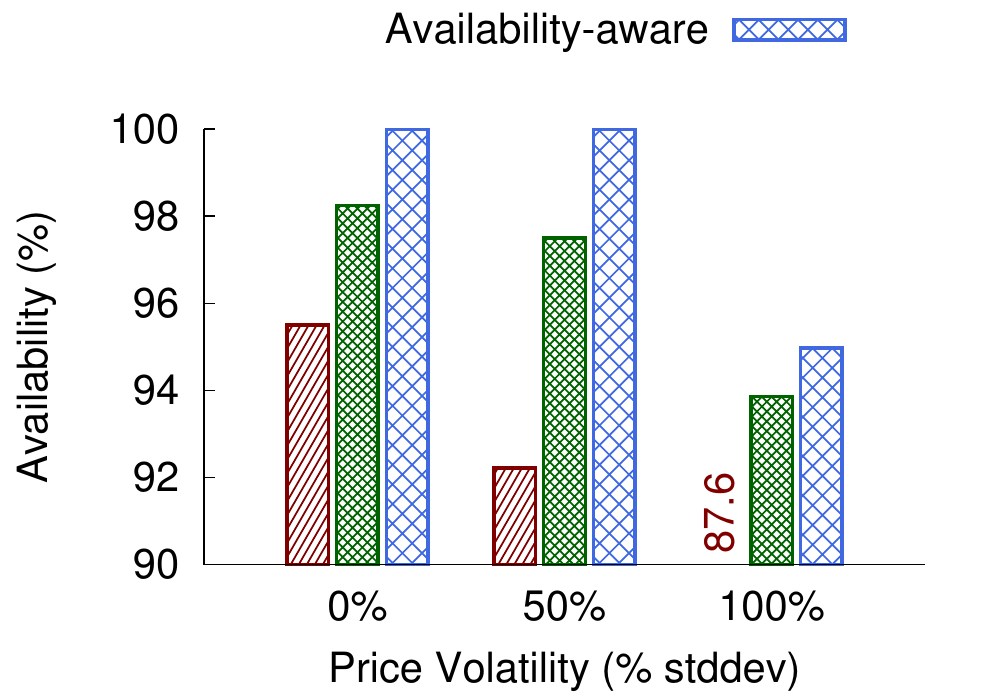}
\caption{Cost-availability tradeoff of different policies as the price volatility changes.  The x-axis is the increase in the standard deviation of prices relative to Figure~\ref{fig:policies-baseline}.}
\label{fig:policies-market-change}
\end{figure}

\noindent {\bf Changing Spot Price Characteristics}. We also vary the spot price characteristics relative to our baseline experiment above.  To do so, we increase the standard deviation of the price changes by the percentage represented on the x-axis of Figure~\ref{fig:policies-market-change}.  A higher percentage indicates an increase in price volatility.  The left graph shows that both the cost-centric and balanced policies improve their cost as market volatility increases, even though the average prices have remained the same.  The price volatility creates an opportunity to benefit from price differentials by migrating. Interestingly, the availability-aware policy, when forced to migrate under more volatile conditions, lowers its costs as well, although not as much as the other two policies. As expected, the right graph shows that availability decreases as the price volatility increases due to the increasing number of migrations.  As before, the balanced index tracking policy lies between the cost-centric and availability-aware policy, having a cost and availability between the two extremes. 

\noindent{\bf Result.} \emph{The balanced index tracking policy, which uses a modified Sharpe ratio that considers both price magnitude and volatility when making migration decisions, balances both cost and availability, achieving a low cost near that of the cost-centric policy and a high availability near that of the availability-aware policy.}

\subsection{Cloud Index Tracking on EC2}
\label{sec:simulation}

We use simulation to evaluate cloud index tracking on real EC2 price traces using production job traces over a long period of time under different policies.  We experiment with three policies and two different types of jobs, as described below.  For these experiments, we use EC2 spot prices from the {\tt us-west-1} region from 2017/03 through 2017/08. We run 3 trials, one for each of the region's AZs ({\tt 1a}, {\tt 1b}, {\tt 1c}), whose index prices are shown in Figure~\ref{fig:zone-indices}, and report the maximum, minimum, and average in the graphs. While each AZ consists of 79 Linux spot VMs, every job considers only the set of spot VMs that meet its minimum resource requirements. We evaluate three policies: a static approach, the cost-centric policy from above, and our balanced index tracking policy.  The static approach selects the optimal spot VM based on its average price and workload characteristics, and runs on the VM until it is revoked, i.e., when its price rises above the on-demand price, and or the job completes. The static approach is similar to EC2's SpotFleet tool, which uses the same policy.  Our cost-centric policy is similar to HotSpot~\cite{hotspot} and other approaches that aggressively migrate to chase low prices.  Finally, our balanced policy uses index tracking. 

We examine the performance of two different applications that are sensitive to downtime and unavailability.  Applications that are not sensitive to unavailability can use the aggressive cost-centric migration policy, as there is no penalty for frequent migrations beyond their associated cost overhead. 

\begin{figure*}[t]
\makebox[1\textwidth][c]{
\centering
\begin{minipage}[c]{1\textwidth}
\hspace{1cm}
\subfloat[Long-running Jobs\label{fig:sim-long-running}]
{\includegraphics[width=0.2\textwidth]{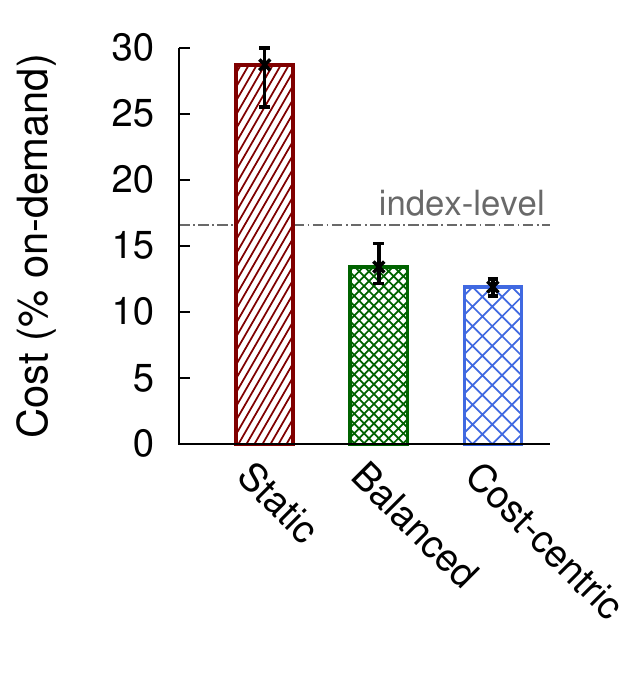}
\includegraphics[width=0.2\textwidth]{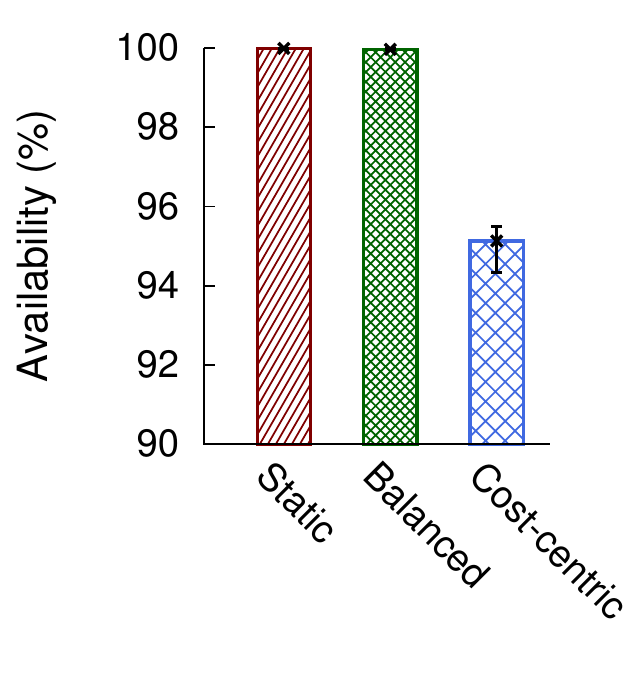}}
\hspace{1cm}
\subfloat[BSP Jobs\label{fig:sim-parallel}]
{\includegraphics[width=0.2\textwidth]{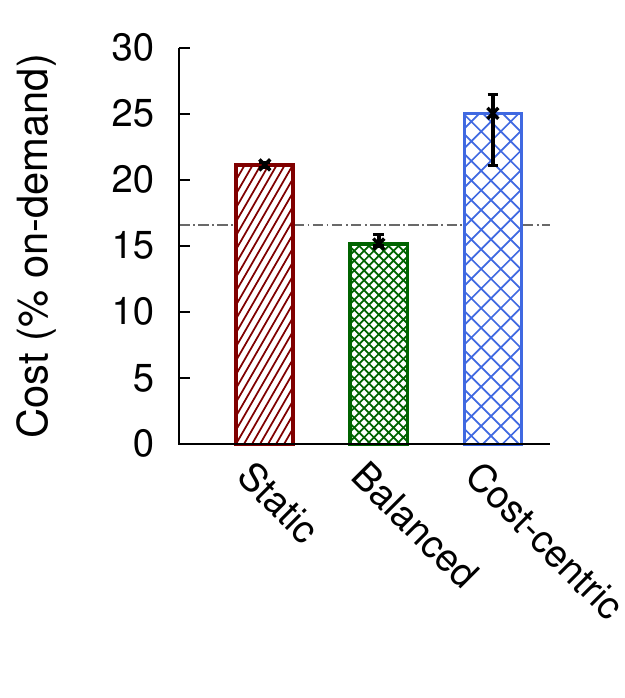}
\includegraphics[width=0.2\textwidth]{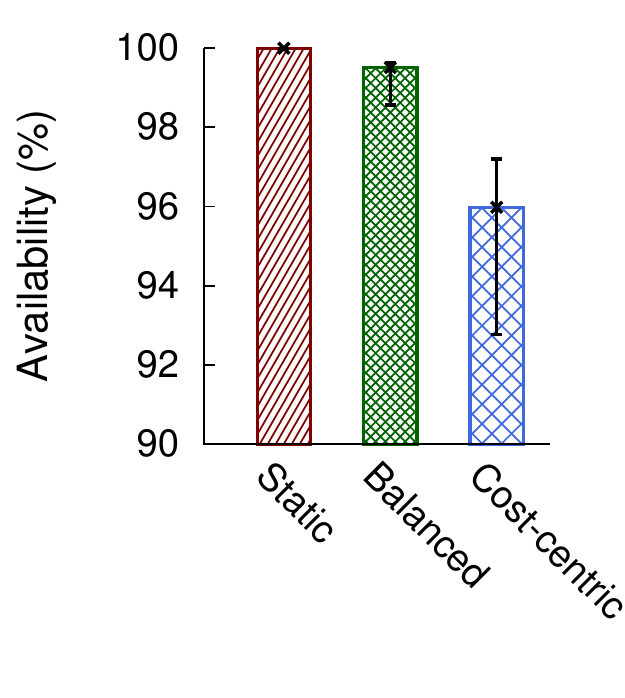}}
\end{minipage}}
\caption{Cost-availability comparison of different policies for both long-running jobs (a) and BSP jobs (b).}\vspace{-0.2cm}
\label{fig:simulations}
\end{figure*}

\noindent {\bf Long-running, Occasionally-Interactive Applications}.  Example applications include data sink servers for IoT devices, cryptocurrency miners, and peer-to-peer file trackers.  These applications can tolerate occasional downtime and restarts due to revocations, but increasing unavailability does decrease the application's utility.  We simulate running these applications for a 6 month period, such that they require a minimum of 2 vCPUs and 10GB memory.  While their performance degrades below this resource capacity, it does not scale up with additional resources. Given this level of resource consumption, we can migrate the application using a stop-and-copy migration with a downtime of 30s.  Finally, the application incurs a downtime of 90s on a VM revocation to acquire and configure a new spot VM, and then restart the application. 

Figure~\ref{fig:sim-long-running} shows both the overall cost and availability of running the application on spot VMs over the 6 month period under each policy.  To establish a baseline, we simulate running the application on the cheapest on-demand VM (the {\tt r4.large}) that meets its resource requirements, and then normalize the execution cost of all policies to this baseline.  The figure shows that all 3 policies are substantially cheaper than using on-demand VMs, ranging from 29\% the cost (for the static policy) to 11\% the cost for the cost-centric policy.  Both the cost-centric and balanced index tracking policy have a cost near that of the index price, represented by the dotted line, with both achieving $>$50\% cost reduction relative to the static approach.  However, the static policy and balanced policy achieve three 9s of availability, while the cost-centric policy has a 95\% availability.  This indicates that aggressively chasing low prices yields little cost reduction relative to index tracking, but significantly reduces availability. Over the 6 month period, the static policy experienced an average of $4.33$ revocations and $0$ migrations; the cost-centric policy experienced no revocations and migrated 4208 times on average; and the balanced index tracking policy experienced 1 revocation and only migrated 24.66 times on average.

\noindent {\bf Bulk Synchronous Parallel Applications}.  Many big data frameworks that run in data centers, such as Hadoop and Spark, follow a bulk synchronous parallel (BSP) programming model.  These applications run parallel tasks on different servers that periodically synchronize with each other.  As a result, the downtime, or performance degradation, for one task can negatively impact other tasks, as all tasks must reach the synchronization point, i.e., barriers, before the application can proceed. We experiment with BSP applications by randomly selecting 1000 jobs from a publicly-available Google cluster trace~\cite{google-cluster-traces}. Internally, each of these jobs comprise multiple worker tasks, ranging from 10 to 500, that execute on separate servers, but periodically synchronize during their execution. The traces report the CPU and memory consumption of each task every 5 minutes.  The job running times vary between 10 and 720 minutes.  We execute each job under the different policies.  Based on the BSP characteristics above, our simulator i) restarts tasks if its VM is revoked, which causes other tasks to pause until the restarted task catches up, ii) pauses all tasks when any task is migrating, since other tasks must wait for the migrating task to synchronize, and iii) requires all tasks to complete before the job finishes.

Figure~\ref{fig:sim-parallel}(right) shows the cost and availability of executing the jobs.  For the cost, we normalize the y-axis to the cost of running all jobs on the cheapest on-demand VM that satisfies their resource requirements.  The graph shows that the balanced index tracking policy not only satisfies the index price, but also exhibits a 30-40\%  lower cost than the other two policies.  In this case, the performance of the cost-centric policy degrades due to asynchronous migrations, i.e., where one or a few workers migrate at different times, which cause a small number of migrating workers to stall other workers, causing them to waste resources waiting until these migrating workers reach the synchronization point.  Since the cost-centric policy operates at the level of a single server, and does not consider synchronization across servers, it performs poorly.  In contrast, our balanced index tracking policy has fewer migrations and they tend to be synchronous, i.e., where all or most workers migrate at the same time.  Even the static policy, which never migrates, performs better than the cost-centric policy since it does not incur the overhead of migration.  The static policy incurred a lower cost relative to the long-running job, since the shorter job lengths reduced the impact of periodic revocations.  The availability follows a similar pattern: the static and balanced policy have 4 and 3 nine's availability, respectively, while the cost-centric policy has 96\% availability. 

\noindent {\bf Result}. \emph{Cloud index tracking yields a predictable cost near that of the index price for both long-running and BSP applications, while also achieving a high availability.}

\section{Related Work}
\label{sec:related-work}

There has been significant prior work on optimizing for variable-priced spot VMs in different contexts.  We summarize this work below and its relationship to cloud index tracking. 

\noindent {\bf Spot Price Modeling and Prediction.} Systems can better optimize for variable-priced VMs if they can model and predict their prices.  As a result, multiple prior works directly or indirectly propose methods of modeling and predicting spot prices~\cite{deconstruct,drafts,drafts2,predict3,predict2,predict1}.  This work differs from our work in that it focuses on modeling and predicting the prices of individual spot VMs, while our work examines spot prices in aggregate using a price index.  We show that these aggregate index prices are more stable and predictable than individual spot VM prices.  While individual pricing models and prediction methods are subject to changing pricing algorithms, our insight that aggregate prices tend to be more stable and predictable than individual prices holds across different pricing algorithms.  Prior work often uses pricing models and predictions to develop bidding strategies~\cite{sigcomm-bid,infocom-bid,no-bid}.  Such bidding strategies are orthogonal to our work, since we assume a fixed bid.  In addition, EC2's most recent pricing algorithm does not use user bids, i.e., their maximum price, as a direct input to their pricing algorithm. 

\noindent {\bf Handling Revocations using Fault-tolerance.} A separate thread of work focuses on optimizing the cost of using spot VMs in the presence of unexpected revocations due to price increases, which impose a performance penalty on applications.  This work generally treats revocations as failures and then optimizes the use of fault-tolerance mechanisms, such as checkpointing and replication, to balance their performance overhead, which increases running time and cost during normal operation, with their performance benefit, which decreases running time and cost if a revocation occurs~\cite{spoton,flint,spotcheck,exosphere}.  Some of this prior work focuses on specific applications, such as Spark~\cite{flint}, and does not operate at the systems level as with cloud index tracking.  While this work also includes spot VM selection policies, which balance the spot price's magnitude and volatility, the policies are static and do not dynamically and transparently migrate applications as prices and resource utilization change over time, as with cloud index tracking. 

\noindent {\bf Cost-centric VM Selection and Migration Policies.}  A number of prior works have implemented intelligent spot VM selection and migration policies.  For example, Smart Spot Instances~\cite{smart-spot} and Supercloud~\cite{supercloud} leverage nested VM migration policies to satisfy various objectives, such as lower cost and access latency, respectively. However, neither approach exposes a similar cost-availability tradeoff, or increases the predictability of costs.  As discussed in \Section\ref{sec:implementation}, our index tracking policy builds on prior work on HotSpot, which implements a cost-centric selection and migration policy that aggressively migrates to the lowest cost spot VM~\cite{hotspot}. As we show, while HotSpot decreases costs, its frequent migrations result in high overhead and unavailability.  As a result, HotSpot is most effective for applications that can tolerate any amount of downtime and unavailability.  In contrast, we focus on applications that can tolerate some downtime and unavailability, but where their utility decreases the more downtime they experience.  We show that cloud index tracking is able to achieve a predictable cost near that of the index price, which is highly stable, while also achieving high availability.  Finally, we have also used a cloud index to inform which regions and AZs have lowest and least volatile prices~\cite{workshop}.  

\section{Conclusion}
\label{sec:discussion}
\label{sec:conclusion}

This paper observes that as we aggregate spot prices for a group of VMs, the aggregated price becomes more stable and predictable, and then discuss the underlying reason for this stability based on current cloud infrastructure and workload characteristics.  We leverage this insight to design a cost-predictive migration policy, which we call cloud index tracking, that automatically migrates to a new spot VM if its current price significantly deviates from the expected index price. Our evaluation shows that cloud index tracking achieves a predictable cost, near that of a cost-centric policy that aggressively migrates to minimize cost, but with much higher availability, similar to that of an availability-aware policy that infrequently migrates to prevent downtime. Importantly, our insights above about cloud index prices are independent of a particular pricing algorithm.  For example, EC2 changed its pricing algorithm in November 2017 to decrease the volatility of individual spot VMs' prices.  However, despite the change in individual spot prices, the magnitude and variance of the index price did not significantly change.  

EC2 altered their pricing algorithm to track long-term changes in the supply of idle capacity, rather changes in both supply and demand, i.e., changes in user bids.  Since, as we discuss in \Section\ref{sec:first-principles}, recent work indicates that idle cloud supply and demand are stable, individual spot prices, while still variable, have become more stable.  However, this stability appears to be due to applications not adapting to changes in their workload or prices, as many enterprise customers treat their cloud infrastructure similar to their on-premises infrastructure, i.e., they mirror their static on-premises infrastructure onto a static set of cloud VMs, which they hold regardless of their utilization~\cite{azure-economics}.  As applications become more sophisticated and cloud native, they will increasingly adapt to changes in their workload and prices, causing the supply of idle capacity to change more frequently, which will require either price volatility to increase, or an imbalanced supply and demand, i.e., periods where capacity is idle or at 100\% resulting in rejected requests. 

EC2 likely changed their pricing algorithm to increase price stability to imitate similar offerings from Google and Microsoft that offer transient servers for a fixed-price.  While this change reduced some of the complexity of using spot VMs and much of their demand-induced volatility, it also eliminated many of their benefits.  By not setting the price based on supply and demand, users have less visibility into when their VMs may be revoked, i.e., they could be revoked when the price is above or below their maximum price, which makes it difficult to optimally configure applications to minimize the performance impact of revocations~\cite{transient-guarantees}.  In addition, users cannot necessarily obtain spot VMs at any time under periods of resource constraint if they are willing to pay a high enough price, i.e., by forcing EC2 to revoke VMs from a lower-paying user.  Since cloud platforms are still growing rapidly, and cloud providers try to provision them to say ahead of the demand growth, these periods of constraint are currently rare.  However, these periods may increase if the growth of cloud platforms stabilize.  

\begin{figure}[t]
\centering
\includegraphics[width=0.45\textwidth]{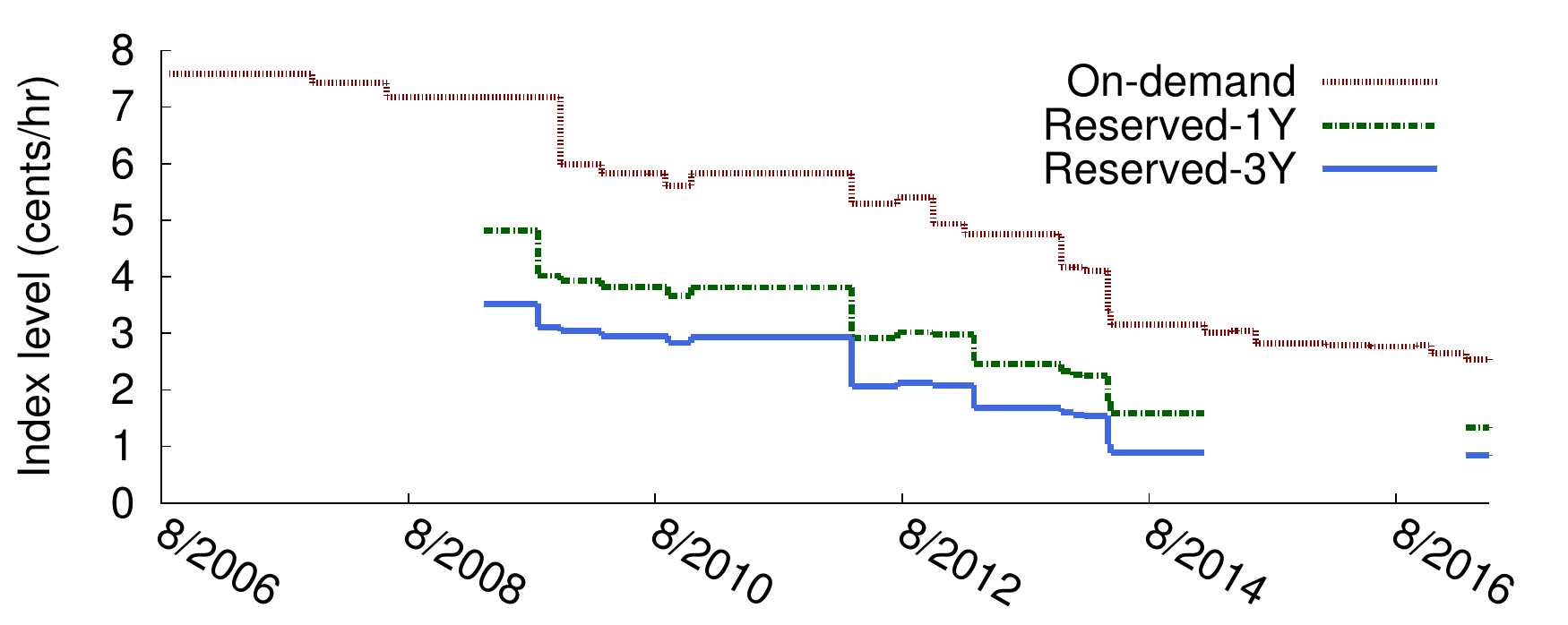}
\caption{Index price of on-demand and reserved VMs in the {\tt us-east-1} region since EC2's inception.}
\label{fig:ec2-history}
\vspace{-0.2cm}
\end{figure}

While we focus on the use of cloud indices for tracking short-term changes in the price of spot VMs, they are also useful as a general tool for tracking cloud prices over much longer time-scales.  This is especially important when making decisions about longer term, multi-year investments, such as reserving VMs for multiple years for an upfront price.  To illustrate, Figure~\ref{fig:ec2-history} shows the index price trajectory of on-demand and reserved VMs in the {\tt us-east-1} region over the last decade.  While reserved VMs are cheaper at any given point in time, they require users to lock in a price for multiple years. However, on-demand VMs experience steady price reductions that reduce the savings relative to reserved VMs. 

\balance
\bibliographystyle{ACM-Reference-Format}

\end{document}